\documentclass [12pt,twoside]{article}           % version LATEX2E
\usepackage[left,modulo]{lineno}
\usepackage{setspace}

\countdef\arxiv=201
\ifnum\arxiv=0 \input cpreb.tex \fi

\ifnum\arxiv=1

\usepackage{epsfig,times,lscape}
\usepackage[usenames]{color}
\usepackage{textcomp}  % Pour faire le symbole ``pour mille''
    \ifnum\count102=1

    \fi

    \ifnum\count102=2
\fi

\newcommand{\nc}{\newcommand}

     \ifnum\count101=1
\nc{\qI}[1]{\section{{#1}}}
\nc{\qA}[1]{\subsection{{#1}}}
\nc{\qun}[1]{\subsubsection{{#1}}}
\nc{\qa}[1]{\paragraph{{#1}}}

\def\qpar{\vskip 2mm plus 0.2mm minus 0.2mm}
\def\qL{\hfill \break}
     \fi 

      \ifnum\count101=2
 \nc{\qI}[1]{\parindent=0mm \vskip 8mm 
{\centerline{\LARGE \color{red}#1}}\vskip 3mm}
\nc{\qA}[1]{\vskip 2.5mm \noindent 
{{\bf\large\color{blue}  #1}} \vskip 1mm \parindent=0mm}
 \nc{\qun}[1]{\vskip 1mm \noindent {\sl #1 }\quad }

\def\qL{\hfill \break}
\def\qpar{\vskip 2mm plus 0.2mm minus 0.2mm}

      \fi
\def\qth{\vrule height 12pt depth 0pt width 0pt}
\def\qtb{\vrule height 0pt depth 5pt width 0pt}

\nc{\qfoot}[1]{\footnote{{#1}}}

\newcommand{\promille}{
  \relax\ifmmode\promillezeichen
        \else\leavevmode\(\mathsurround=0pt\promillezeichen\)\fi}
\newcommand{\promillezeichen}{%
  \kern-.05em%
  \raise.5ex\hbox{\the\scriptfont0 0}%
  \kern-.15em/\kern-.15em%
  \lower.25ex\hbox{\the\scriptfont0 00}}

      \ifnum\count101=1
\def\qbu{\hfill \par \hskip 6mm $ \bullet $ \hskip 2mm}
\def\qee#1{\hfill \par \hskip 6mm (#1) \hskip 2 mm}
      \fi
      \ifnum\count101=2
\def\qbu{\hfill \par \hskip 4mm $ \bullet $ \hskip 2mm}
\def\qee#1{\hfill \par \hskip 4mm (#1) \hskip 2 mm}
      \fi

\def\qparr{ \vskip 1.0mm plus 0.2mm minus 0.2mm \hangindent=10mm
\hangafter=1}

     \ifnum\count101=1 
 \def\qdec#1{\parindent=0mm\par {\leftskip=2cm {#1} \par}}
     \fi
     \ifnum\count101=2
  \def\qdec#1{\parindent=0mm \par {\leftskip=1cm {#1} \par}}
  
  \def\qcitb#1{\noindent \hbox to 102mm{\hfill \small #1} \vskip 1mm}
      \fi

 \def\qpages#1{\count102=0{\loop\advance\count102 by 1
 \null \vfill\eject \ifnum\count102<#1 \repeat}}

\def\qth{\vrule height 12pt depth 0pt width 0pt}
\def\qtb{\vrule height 0pt depth 5pt width 0pt}

\def\qv{\vskip 0.1mm plus 0.05mm minus 0.05mm}
\def\qhu{\hskip 0.6mm}
\def\qhv{\hskip 3mm}

\def\qhw{\hskip 1.5mm}
\def\qleg#1#2#3{\noindent {\bf \small #1\qhw}{\small #2\qhw}{\it \small #3}\qv }
\fi
\begin{document}
\thispagestyle{empty}

% --------------------------------------------------------------------

      % Hauts de pages et numerotation

          % Remarque: sans le \protect --> message d'erreur (ordre fragile)
\markboth{{\sl \hfill  \hfill \protect\phantom{3}}}
        {{\protect\phantom{3}\sl \hfill  \hfill}}

% -------------------------------------------------------------------
\color{yellow} 
%\hrule height 20mm depth 10mm width 170mm 
\hrule height 10mm depth 10mm width 170mm 
\color{black}

\vskip -16mm

\centerline{\bf \Large The future of US-China relations:}
\vskip 3mm
\centerline{\bf \Large a scientific investigation}
\vskip 15mm

\centerline{\large 
Belal E. Baaquie$ ^1 $,
Peter Richmond$ ^2 $, 
Bertrand M. Roehner$ ^3 $ and
Qing-hai Wang$ ^4 $}

\vskip 10mm
\large

{\bf \color{red} Abstract}\qL
In earlier centuries kings and governments
employed astrologists to help them take the
best decisions.
Present-day governments no longer employ astrologists
but still  have no clear analytical tool
to replace them.
Over the past two decades we have developed a methodology
for the scientific investigation of recurrent historical
events. It consists in two steps.
(i) Identification and comparison of
historical episodes driven by a common mechanism.
(ii) Under the reasonable assumption that what has
happened several times in the past is likely to happen again,
one then derives testable predictions. This of course is
nothing other than the protocol used in experimental
science when exploring new phenomena.
We believe such a tool
can give decision makers much better insight.\qL
In the present paper we illustrate this analysis by
considering challenges, that span more than a century, to US hegemony
in the Pacific.
The outcomes suggest that it is only through the
sidelining of one of the contenders that the confrontation will
end.
At the time of writing (late 2018) early evidence of this confrontation
is already visible at three levels.
(i) Growing US concerns for domestic security
that are leading to a new form of McCarthyism.
(ii) Political instability due to China-US
polarization in several Asian countries
as well as in the countries participating in
the ``Belt and Road Initiative’'.
(iii) Tension and sanctions in procurement and trade.

\vskip 2mm
%\centerline{\it \small Provisional. Comments are welcome.}
\centerline{\it \small Version of 26 January 2019}
\vskip 2mm

{\small Key-words: International relations, China, United States,
historical events, recurrent, prediction.}

\vskip 5mm

{\normalsize

1: INCEIF (International Centre for Education in Islamic Finance), The
Global University of Islamic Finance, Lorong Universiti A, 59100,
Kuala Lumpur, Malaysia. Email: belalbaaquie@gmail.com \qL
2: School of Physics, Trinity College Dublin, Ireland.
Email: peter\_richmond@ymail.com \qL
3: Institute for Theoretical and High Energy Physics (LPTHE),
Pierre and  Marie Curie campus, Sorbonne University, Paris, France.
CNRS (National Center for Scientific Research).\qL
Email: roehner@lpthe.jussieu.fr.\qL
4: Physics Department, National University of Singapore.
Email: qhwang@nus.edu.sg
}

\vfill\eject

\null
\vskip 5mm

In this paper we address two questions.
\qbu During past decades successive Chinese leaders have
repeatedly declared that their objective is a win-win relation
and shared leadership with the United States, but is the
US side ready to share world hegemony?
\qbu As we shall see, the answer appears to be no and we then ask the second
question namely: what form does the confrontation take?
\qpar

The distinctive feature of the approach is that it is not
based on a discussion of present circumstances but on
regularities identified within a number of similar events
that have previously occurred.

\qI{Introduction}

The approach taken by historians can hardly
be considered to be science (in the sense of
being able to produce testable predictions). Furthermore many historians even
disagree with the very notion
that history can be made into a science.
\qpar

Here we take the contrary view, acknowledging that
the defining condition of any scientific investigation
is {\it replicability}.
For example, if the falls of apples and rain drops have
common characteristics we may be able to make predictions relating to
the fall of
hailstones?
\qpar

Our methodology follows the approach that experimental
physics uses to explore new phenomena. It is very much in the
spirit of the methodological guidelines defined by 
the French sociologist Emile
Durkheim in his book%
\qfoot{Indeed, Durkheim recommends
studying social phenomena as if they were ``things''
(``comme des choses'' in the French text).}
entitled ``The rules of
sociological method'' (Durkheim 1895).
It is perhaps not suprising that
Durkheim's methodology is similar to that of experimental physics since
at the end of the 19th century physics
was seen as {\it the} compelling approach in science.
\qpar

To study recurrent events
the decisive requirement is replicability simply because
otherwise
nothing but noise is observed.
But note that even in physics replicability is
never perfect%
\qfoot{This point is discussed further with reference to a specific example in
Appendix C. A broader analysis of the
key importance of the signal to noise ratio can be found
in Roehner (2007) in a chapter entitled ``The battle
against noise in the social sciences''.}%
.
When  we consider
a cluster of similar historical events we must be prepared to see
fluctuations that are substantially
larger than in most physical observations, and it is necessary to select
such events in such a way that the dispersion is minimised.

\qA{Precedents in the move toward a scientific approach}

In order to give credence to the possibility of
establishing a ``science of recurrent historical events''
(which in what follows will be called ``analytical history''),
it is useful to describe briefly how the transition to
science status was achieved in two other fields, namely
astronomy and medicine. Such transitions are summarized in
Table 1.
\qpar

It was to provide advice on key-decisions
that astrology was developed. This sought to ``rationalize'' the activity of
numerous oracles and shamans who were considered intermediates
between humans and the world of spirits and gods.
From China and Mesopotamia, to
Greece and western Europe the practice of astrology
was wide\-spread. We call it a neo-science because it has
some characteristics of a real science.
It is based on the relative positions of celestial objects
which in turn necessitated accurate observations of the positions
of stars and planets. In the evolutionary perspective schematized
in Table 1, we use the expression neo-science
(one could also call it pre-science) rather than pseudo-science
because the later conveys a derogatory meaning. Initially,
astrology marked progress with respect to
divination at least in the kind of data
that it was handling.
\qpar

%
%%-----------------------------------------------
\begin{table}[htb]

\small
\centerline{\bf Table 1: Transitions from
divination stage to neo-science and to science.}

\vskip 5mm
\hrule
\vskip 0.7mm
\hrule
\vskip 0.5mm
$$ \matrix{
\qtb
\hbox{Divination stage}\hfill & \hbox{Neo-scientific factual stage} \hfill &
\hbox{\color{blue} Scientific stage}\hfill \cr
\noalign{\hrule}
\qth
\hbox{\it \quad Transitions already completed}\hfill & \hbox{} \hfill &
\hbox{\color{blue} }\hfill \cr
\hbox{Prophecies by oracles}\hfill & \hbox{Astrology} \hfill &
\hbox{\color{blue} Astronomy}\hfill \cr
\hbox{Shamanism (vital spirit, soul)}\hfill & \hbox{Humoralism,
 bloodletting} \hfill &
\hbox{\color{blue} Medicine}\hfill \cr
\hbox{}\hfill & \hbox{} \hfill &  \hbox{}\hfill \cr
\hbox{\it \quad Transition under way}\hfill &
\hbox{} \hfill &
\hbox{\color{blue} \it }\hfill \cr
\qtb
\hbox{Founding myths, legends}\hfill &
\hbox{Factual historical narratives} \hfill &
\hbox{\color{blue} \it Analytical history}\hfill \cr
\noalign{\hrule}
} $$
\vskip 1.5mm
Notes:\qL
``Humoralism'' was based on four, so called, humors whose
equilibrium was deemed to be necessary for a healthy life.
Introduced by Hippocrates
(460 to 370BC), this view was accepted for over 2,000 years.
The neo-scientific stages defined in the 
second column of the table differ from those in the first in two main ways:
(i) They make little or no reference to gods or supernatural notions.
(ii) They were formalized
into well defined protocols.
Thus, following Galen (129-200AD) the location and volume
of blood lettings were defined by a
complex set of rules based on disease, age, season and weather.\qL
The transitions between stages were gradual
rather than sudden.
For instance, although Galen was wrong on humoralism,
many of his anatomical observations were
correct and proved useful for centuries. 
Incidentally, even today, treatments
without real
medical justification continue to be performed, as explained in
Baugh et al. (2011) for the case of
tonsillectomy, i.e. removal of the tonsils on both sides of the
pharynx (for more details
see the Wikipedia section entitled ``Tonsillectomy industry'').
%Sources:
\vskip 2mm
\hrule
\vskip 0.7mm
\hrule
\end{table}
%%-----------------------------------------------
In the history of mankind
the transition from astrology to astronomy took
place only a few times: first in Greece, 
before the Roman invasion,
then around 900 in the Islamic world and finally 
in Western Europe around 1600.
It is also significant here to note that both the Danish astronomer Tycho Brahe
(1546-1601) and Johannes Kepler (1571-1630) were
astrologists at the service of the German Emperor Rudolf II.
The present-day consensus is that
the astrologists' claim of predicting terrestrial
events based on celestial
observations is not justified%
\qfoot{Yet, very few really scientific tests have been
conducted.
One is described by Shawn Carlson in  ``Nature'' (1985).
The paper concludes that predictions based
on so-called natal charts are no better than pure chance predictions
but also emphasizes that such tests involves many methodological
difficulties. Would experimental tests on animal
populations not offer better tests of astrological
predictions? If
planets have an influence on humans, then they must
also have an influence on animals.}%
.
Describing astrology as a ``pseudo-science'' emphasizes the point
that its claims are not justified.
\qpar

As we have seen
the transition from astrology to astronomy was gradual. So was
also the transition from alchemy to chemistry.
The same observation holds for medicine. 
Even though, during the neo-scientific stage, 
weird procedures such as relief by bloodletting 
were followed, significant
contributions were made to anatomy and herbal medicine.
\qpar

Returning to history, although
we are currently in the neo-scientific stage, the transition
to the scientific stage does not seem welcome or desirable
by mainstream historians. That is perhaps why physicists
have a role to play. When astrophysicists want to study
a law of neutron stars, quite naturally they do not
study just one neutron star but as many as possible.
Our approach follows the same methodology.
\qpar

Incidentally, the transition from the divination stage to
the factual stage was not only gradual but also
marked by steps backward. Thus, whereas the accounts written by
Herodotus (circa 484 to circa 425 BC) or
Thucydides (460 to 400 BC) were fairly factual, 
when one reads
contemporary accounts of the ``Great London Fire'' of 1666
one learns more about the sins of people and the intentions
of God than about the real circumstances of the conflagration.
\qpar

Presently, although
there is no science of historical events,
some sub-fields of history, e.g. demography,
already function as sciences. The main obstacle to
the development of analytical history is probably the fact
that most historians consider
historical events to be unique and therefore inappropriate for
scientific analysis. We return to this point in Appendix C.

\qA{Outline of the investigation of challenges to US hegemony}

The investigation will proceed through the following steps
\qbu First, we define the mechanism on which we shall focus.
\qbu Secondly, we identify historical episodes showcasing
this mechanism.
\qbu Thirdly, we compare their outcomes and consider to what extent
they help predict the future of China-US relations.
\qbu Finally, we examine recent events that may confirm
and foreshadow impending confrontation.
\qpar

Because not all readers may need or even wish to know the details
of our arguments explanations that rely on historical
facts and accounts are in Appendices A and B.
However we emphasise that such
``details'' are often of crucial importance.
For instance, our belief
(used in Table 2)
that in 1900 the United States had already an hegemonic
position in the Pacific must be substantiated and we do this by
showing that at that time China had lost its power, Japan was
still in a development stage and Britain had
accepted US ascendancy in return for a free hand in
Hong Kong and Malaysia.

\qI{Investigation of challenges to US hegemony in the Pacific}

\qA{Definition of the mechanism}

The question investigated in this paper can be stated as follows.

\qdec{\it {\color{blue} 
Class of events: reactions to actions challenging hegemony.}\quad
We consider situations in which a
country (or company or organization), $ A $,
holds a dominant position but is
challenged by another country, $ B $. What will happen?.}
\qpar

To say that a historical episode involving two countries $ A,B $
belongs to the class defined above one must
show two things:
\qee{i} That $ A $  holds an hegemonic position
in a given area, meaning that  $ A $ is more powerful 
in terms of GDP, technology,
armed forces than all other countries in this area.
\qee{ii} That, at least in the minds of the leaders of $ A $,
the action of $ B $ is challenging this position%
\qfoot{Because it relies not only on objective
facts but also on their perception, this notion
is not always easy to define precisely.}%
.
\qpar

Sometimes it happens that a country has to give up
hegemony just as a result of general circumstances.
This is how hegemony was transfered from Britain to the
United States. The reason is easy to understand.
Basically the two world wars made Europe weaker and
the United States stronger. In 1939, the US GDP
was already 3.4 times larger than the GDP of the UK.
In 1950, the ratio was 7.9%
\qfoot{The data are as follows (Liesner 1989, p.20-21,54-55,74-75):\qL
1939: UK GDP=5.96 billion pounds, US GDP=bn \$91.3, 1 pound=\$4.46 \qL
1950: UK GDP=13.0 bn pounds, US GDP=\$288 bn, 1 pound=\$2.80.}%
.
In addition, the UK was heavily indebted to the US.
Thus, the transfer of leadership could hardly be questioned.
This case illustrates the fact that, as in any competition,
the contest is of interest only when the two contenders
are approximately of same strength.
\qpar

In the present paper we analyze the dominant
position of the United States in the Pacific.

\qA{US hegemony in the Pacific: general view}

\qun{Why do we limit the study to the Pacific?}
\qL
Historically there have been many cases marked by recurrent
conflicts between regional contenders.
For instance, one can mention
the series of conflicts between Venice and Genoa from 1250 to 1400,
the four Anglo-Dutch wars fought on sea from 1650 to 1780,
the many wars between Russia and the Ottoman empire. However,
all these cases were rather imperfect examples of challenges
to hegemony. They were rather conflicts 
between two powers of similar strength.\qL
The position of the United States in the Pacific 
is a much clearer case of hegemony.
\qpar

Nowadays (in 2018) what is at stake is not only
hegemony in the Pacific rim but world hegemony.
However, the question of world hegemony came on the
table only after the demise of the Soviet Union as a
global contender. In contrast US hegemony claims 
in the Pacific are much older. 
\qpar

On 2 June 1954 at a White House
Security Conference, President Eisenhower declared:
``We have got to keep the Pacific as an American lake''.

%
%%-----------------------------------------------
%%%% Pacific rim
\begin{figure}[htb]
\centerline{\psfig{width=17cm,figure=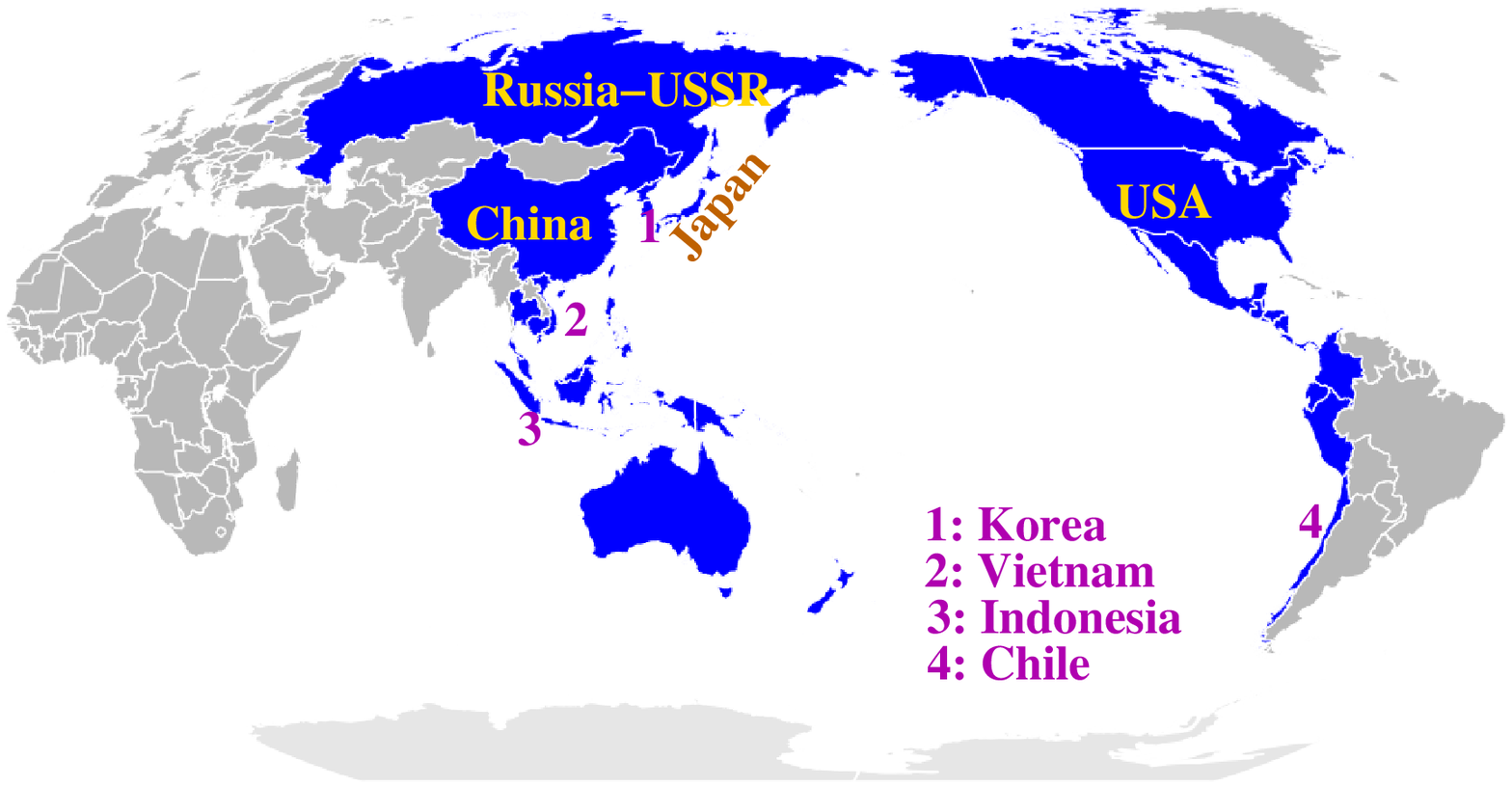}}
\qleg{Fig.\qhu 1\qhv Rim of the Pacific}
{A number of historical circumstances (explained in the text)
explain that the United States gained an hegemonic
position in the Pacific as early as the end of the 19th
century through the Spanish American war of 1898.}
{}
\end{figure}
%-------------------------------------------------

In contrast to a sea, a lake is most often included
within the territory of a country. Naturally, here
this sentence does not mean that all the countries
around the Pacific should be American but
it implies that they should be part of the US
zone of influence%
\qfoot{Before World War II the 
notion of ``zone of influence'' was well accepted and
duly used in diplomatic language. Nowadays, in contrast,
it has been replaced by the delusion that UN membership
makes even the smallest countries (e.g. ``Saint Kitts and Nevis''
or Kiribati) fully independent in all and every respects.}
that is to say should be neither contenders nor opponents.
\qpar

Actually, we will show that the US hegemony in the Pacific
did not start after World War II but in fact 
several decades before, around 1900.
Therefore, by restricting our study to the Pacific rim (see Fig. 1)
we can draw on more cases than by considering world hegemony
after 1990.
\qpar

Before closing this subsection we wish to draw attention
on an important distinction.

\qun{Similar historical episodes or similar mechanisms?}
\qL
The starting point of our comparative analysis
must be a {\it mechanism}, it 
cannot merely be an historical fact or episode.
\qpar

For instance, in China there have been successive dynasties
and various mechanisms could be at work in their falls
(e.g. see Ferguson 2010)
For instance, some collapses may be due to military factors,
others to a contraction of tax income or some other factors.
In other words, a first step must be to make sure that the
falls under consideration belong to the same
category.  Incidentally,
one faces exactly the same problem in medicine in the
sense that a comparative analysis of symptoms will be
meaningfull only once the disease has been clearly defined.
For instance,
there must be a clear distinction
between tuberculosis and influenza for, although the two
diseases attack the lungs, they affect them in different ways. 
Similarly, an analysis of
dynasty falls will be all the more fruitful that the
set under consideration will be more homogeneous.

\qA{Historical realizations of this mechanism}

As already said, for each of the cases included in Table 2 we must 
examine two points. (i) Whether there was indeed an hegemonic position.
(ii) Whether there was a real challenge

\qun{Hegemonic position}
\qL
Everybody would certainly agree that after Word War II 
the United States enjoyed a hegemonic position 
in the Pacific. However, 
case number 1 of Table 2 assumes that the US
had already such a position around 1900 which
is less obvious and requires some explanations.
In Appendix A we explain the factors
which account for the fact that US dominance started at the
end of the 19th century.

\qun{Actions perceived as challenges to hegemony}
\qL
Once we agree that as early as 1900 the United States 
had a dominant position in the Pacific, we need to understand why 
the actions of other countries listed in Table 2
were perceived by the US as challenges to this position.
\qpar

For most of the cases (namely Pacific War, Korea and Vietnam War)
this is fairly clear
but there are two cases, namely Russia and Indonesia,
for which additional 
explanations may be helpful. A historical  discussion 
can be found in Appendix B.
\qpar

In the last three decades of the 19th century the situation in
the Pacific can be described as follows.
\qbu China, once the dominant power in East Asia, had become very
weak. The two Opium Wars of 1842 and 1856--1860 had shown
that a relatively small western army supported by the guns
of a naval squadron could defeat fairly easily 
the armies of the Qing Empire. After this demonstration, China
became a prey for western countries just as the decaying Ottoman
Empire was in the Mediterranean area.
\qbu Through its success in the First Sino-Japanese war (1894--1895),
Japan had shown that it was on the way of replacing China
as a regional power. But, as shown in Appendix A, economically
it was still far behind the US and even substantially 
behind Russia. In 1900 its total trade was 2.6 times smaller
than the trade of Russia%
\qfoot{The numbers were 1,340 million rubles for Russia
and 500 milion yen for Japan; the ruble:yen exchange was almost 1:1
(Mitchell 1978,p.306).}%
.
\qbu Until 1898 Spain was present in the Philippines but
its quick defeat in the Spanish American War showed it out.
\qbu It is true that other western powers, particularly Britain,
France and Germany were interested in occupying Pacific
islands. However, the rapid acceptance by Britain of the
US annexation of Hawaii showed that British interests were
confined to the Indian ocean and that it did not wish to
challenge US influence in the Pacific.
\qpar

In other words, the only real possible contender was Russia.

\qun{US hostility toward Russian expansion}

In Appendix B we explain some of the steps in the Russian
expansion
toward Manchuria, Korea and the shores of the Pacific. 
\qpar

Actually,
apart from the geopolitical situation, there was another
factor which amplified the US {\it perception} of Russia
as an unpleasant contender. It was the fact that between
1880 and 1905 there were in Russia
recurrent waves of anti-Jewish pogroms
which attracted great international attention especially
in the United States. These events
amplified US hostility toward Russia in a way somewhat similar
to present day US perception of China as a threat is
amplified because China is not only a contender but is ruled
by the Communist party.
\qpar

The only other cases to require
some more explanations (given in Appendix B)
are Indonesia and Chile.

\qI{Analysis of a cluster of challenges}

Now we put together the information collected and
described previously for the different cases of challenges.
This leads to the cluster of challenges displayed in Table 2.
What conclusions can one draw?

%
%%-----------------------------------------------
\begin{table}[htb]

\centerline{\bf  \color{blue} Table 2: Cluster of
recurrent challenges to US hegemony in the Pacific}

\vskip 3mm
\hrule
\vskip 0.8mm
\hrule
\vskip 1mm

\color{black} 
\small

$$ \matrix{
 & \hbox{Beginning} \hfill & \hbox{\color{red}Challenger} \hfill & \hbox{Country} \hfill & \hbox{\quad} &
\hbox{Country} \hfill & \hbox{\color{blue}Conflict} \hfill &  \hbox{Direct US}\cr
 & \hbox{of} \hfill & \hbox{} \hfill & \hbox{acting for} \hfill & \hbox{\quad} &
\hbox{acting for} \hfill & \hbox{} \hfill &  \hbox{action}\cr
\qtb
 & \hbox{challenge} \hfill & \hbox{} \hfill & \hbox{challenger} \hfill & \hbox{\quad} &
\hbox{US} \hfill & \hbox{} \hfill &  \hbox{(no/yes)}\cr
\noalign{\hrule}
\qth
1 & 1890 \hfill & \hbox{\color{red}Russia} \hfill & \hbox{None} \hfill
& \hbox{} &
\hbox{Japan} \hfill & \hbox{\color{blue}Russo-Japan War: 1904-1905} \hfill &  \hbox{no}\cr
 &  &  & & & & & \cr
2 & 1938 \hfill & \hbox{\color{red}Japan} \hfill & \hbox{None} \hfill
& \hbox{} &
\hbox{Several} \hfill & \hbox{\color{blue}Pacific War: 1941-1945} \hfill &  \hbox{\bf yes}\cr
 &  &  & & & & & \cr
3 & 1949 \hfill & \hbox{\color{red}USSR} \hfill & \hbox{North Korea} \hfill & \hbox{} &
\hbox{Several} \hfill & \hbox{\color{blue}Korea War: 1950-1953} \hfill &  \hbox{\bf yes}\cr
4 & 1948 \hfill & \hbox{\color{red}USSR} \hfill & \hbox{North Vietnam} \hfill & \hbox{} &
\hbox{France} \hfill & \hbox{\color{blue}Indochina War: 1949-1954} \hfill &  \hbox{no}\cr
5 & 1962 \hfill & \hbox{\color{red}USSR} \hfill & \hbox{North Vietnam} \hfill & \hbox{} &
\hbox{Several} \hfill & \hbox{\color{blue}Vietnam War: 1960-1978}
\hfill &  \hbox{\bf yes}\cr
6 & 1957 \hfill & \hbox{\color{red}USSR} \hfill & \hbox{Indonesia}
\hfill & \hbox{} &
\hbox{Indon.army} \hfill & \hbox{\color{blue} Mass murders: 1965-1966} \hfill &  \hbox{no}\cr
7 & 1970 \hfill & \hbox{\color{red}USSR} \hfill & \hbox{Chile} \hfill
& \hbox{} &
\hbox{Chile army} \hfill & \hbox{\color{blue} Near civil war: 1970-1973} \hfill &  \hbox{no}\cr
8 & 1979 \hfill & \hbox{\color{red}USSR} \hfill & \hbox{Afghanistan}
\hfill & \hbox{} &
\hbox{Mujahideen} \hfill & \hbox{\color{blue}1st Afghan War: 1979-1988} \hfill &  \hbox{no}\cr
 &  &  & & & & & \cr
\qtb
9 & 2010 \hfill & \hbox{\color{red}China} \hfill & \hbox{?} \hfill &
\hbox{} &
\hbox{Several} \hfill & \hbox{\color{blue}?} \hfill &  \hbox{?}\cr
\noalign{\hrule}
} $$
\vskip 0.5mm
Notes: \qL
Because cases 1-8 all led to the elimination of the challenge,
they suggest
that the emergence of a contender will simply not be
tolerated. This is even more likely when the contender is a major power
occupying a vast area of the Pacific rim as is China.
As this has occurred several times in the past, it is likely to be
repeated again. 
Despite the interrogation marks left in the row for China,
one can draw the
conclusion that {\it the US-China tension
will not abate but (most probably) will rather wax.}\qL
Below we give a number of additional comments.\qL
$ \bullet $ Direct US action means that US troops were involved.\qL
$ \bullet $ ``Mujahideen'' does not refer to a country but to Muslim
people waging
a ``jihad'' that is to say warfare against unbelievers.
Although Afghanistan is not on the Pacific rim, the fact that it has
been fiercely
disputed for almost four decades suggests that, for some reason, it is
considered
of high strategic value. \qL
$ \bullet $ France waged the ``Indochina War'' on behalf of the US in the sense 
that, along with weapons and funding, the US side gave strategic and
tactical advice.\qL
$ \bullet $ It can be noted that every time the US took directly part
in a war it had several allies; for instance,
Australia, Britain, China,
New Zealand and the Netherlands in the war against Japan.\qL
$ \bullet $ Note that the
Indochina and Vietnam wars were not challenges to US hegemony
by themselves but because of the ``domino effect'' theory which
speculated that if a state came under Communist control, 
then all surrounding countries would also fall one by one.
Despite eventually being proved wrong (indeed after the Communist victory 
in Vietnam no other country changed side),
this theory was used as a justification for massive US intervention
by successive US administrations.\qL
$ \bullet $ The case of Chile was different from the others in the
sense that the two sides
belonged to the Chilean society which is why the confrontation took the form
of a low intensity civil war. \qL
$ \bullet $ In this paper, we do not include the trade war between Japan and US
during 1980s-90s because that conflict remained constrained in the
economic sector.
Japan did not challenge the US politically or militarily.
\qL
{\it Sources: \qL
https://en.wikipedia.org/wiki/Communist\_Party\_of\_Indonesia;
Di et al. (2017), Chapter 5: ``The Pacific as 
an American lake''.}
\vskip 1mm
\hrule
\vskip 0.7mm
\hrule
\end{table}
%------------------------------------------
%

The challenges listed in Table 2 are not all of
same kind. Some like Japan, Korea, Vietnam or Afghanistan were 
military challenges from the very start, others were
ideological and political challenges.
It is the case of Russia which
comes closest to the present situation of China as an 
expanding economic power. On the other hand
the struggle for world
hegemony is best described by cases 3-5 corresponding to
indirect military confrontations with the Soviet Union.

\qA{Overall regularities}

In the regularities displayed by Table 2
we wish first of all to list the regularities deriving
from cases 1-8. These are valid for any challenger
irrespective of its specific identity. Then in the next
subsection we give some comments which apply specifically
to China.
\qee{1} The main conclusion is that 
in the past no challenges have been tolerated. All possible
challengers were targeted and countered.
\qee{2} The challengers were defeated in all but one case,
namely Vietnam.
Note that regularity (1) holds even in this case. It is only
after a bitter struggle and in the face of domestic protests
that defeat was accepted.
\qee{3} In all cases, the United States first relied on
local forces supported by US military advisers and delivery of
weapons. US troops were sent in only when local forces 
turned out unable to defeat the challenger.
\qee{4} Whenever the US intervened directly it was always
together with several allies. In the Pacific War it had
5 major allies namely Britain, China, France, the Netherlands,
and the Soviet Union. It is in the second Vietnam War that the number
of major allies was smallest: neither Britain nor France
took part.
\qee{5} In all cases where it occurred, direct US interventions
were in response to an aggression. This is of course clear
for cases 2,3 and 8. The US administration wanted 
a similar pretext for Vietnam.
Before sending reinforcements 
the Department of Defense wished to
show that it was responding to an aggression.
This led to the two so-called ``USS Maddox incidents''
in the Gulf of Tonkin (2 and 4 August 1964). 
Later on, it was
recognized by US authorities that the second incident did not
take place. Even the first was quite unimpressive in the
sense that not a single American was injured%
\qfoot{Moreover, the US government did not disclose that the
incident of 2 August was preceded on 30,31 July by attacks
against two North Vietnamese offshore islands.}%
.
Within hours
President Johnson ordered retaliatory
air strikes against North Vietnam and announced them
to the nation in a TV address broadcast on all channels
at 23:30 on 4 August.
In response, on 7 August 1964, the US Congress passed the 
``Gulf of Tonkin Resolution'' which 
authorized President Johnson
to use US armed forces to assist any  
state in  Southeast Asia requesting assistance in defense of its
freedom. The vote was  476-0  in the ``House of Representatives''
and 88-2 in the Senate.\qL
On 9 August 1964, China said that the US was trying to create
a United Nations force to turn Vietnam into a second Korea
and pledged to aid North Vietnam. A massive anti-US
demonstration took place in Beijing. Less than two months later
China tested its first atomic bomb.
\qpar

Readers may wonder why we gave a fairly detailed account of this
episode. The reason is simple.  
One expects that whenever the US will decide
to escalate the military confrontation there will be a similar
attempt to present it as a response to an aggression. 
For instance an incident between a Chinese warship and
an Australian, Filipino or Vietnamese vessel would 
allow the US media to portray the 
US Navy as coming to the rescue of a weaker country..

\qA{Time schedule}

Can one say that the confrontations start with a trade war?
Not necessarily. The case of Japan shows 
that US-Japanese trade remained active until
a few months before the war. In 1931--1935 the
average annual US exports to Japan was \$169 million
whereas it was 238 in 1940. These exports represented only
about 3.5\% of the Japanese GDP. As a matter of
comparison, in 2017 Chinese exports to the US
represented about 4\% of the GDP of China. So, one
cannot say that economic interdependance 
is strong enough to prevent a confrontation.
\qL
If needed, this case shows that economic and
political rivalry are definitely two different things.
Economics is about production and consomption whereas 
politics is about power and hegemony.

\qpar

In cases 1-8 the average time interval between
the start of the challenge and its elimination was
about 8 years%
\qfoot{Russia: 15, Japan: 4, Korea: 4, Vietnam(1): 6,
Vietnam (2): 16, Indonesia: 8, Chile: 3, 
Afghanistan: 9. The average is: $ m=8.1\pm 3.6 $
(where 3.6 corresponds to a confidence level of 0.95).
As in several cases the start of the challenge 
is not well defined, this estimate
should be seen as a mere order of magnitude.}%
.
\qpar

What can one say about the duration of economic and financial
challenges? For instance, in Di et al. (2017)
there is an estimate of how long it may take for the
renminbi to replace the dollar as the dominant currency
in the reserves of central banks. This estimate is
based on only one case namely the replacement of the
British pound by the US dollar. 
The time constant was of the
order of 50 years; for the RMB-US\$ replacement it may
well take longer because the dollar holds a much stronger
position than the pound had back in the 19th century.
\qpar

It is true that the composition of
central bank reserves is not the only 
possible criterion. However, it may reflect the weight
of a currency in global {\it financial} transactions.
As a trading currency the weight of the renminbi
may increase faster but in fact in present-day 
market organization
trade and financial transactions are closely interconnected,
particularly through the mechanisms of
currency exchange, trade insurance and
debt management. In short, it does not seem
that the RMB is a short- or medium-term 
threat for the domination of the US dollar
However, what matters is not the
threat itself but rather its perception in the 
minds of American leaders.
\qpar

We now give some comments which are specific to the case
of China.

\qA{The China-US honeymoon era}

In order to understand why awareness of nascent US
hostility took decades to spread in the Chinese
public and in the Communist Party 
it is necessary to remember the situation
prevailing in the period 1980-2018. 
\qpar
Numerous statements by Chinese leaders reveal
that their vision of US-China relations was a kind
of condominium in which the US and China would share
world leadership. It is true that between 1980 and 1990
China was considered by the US as a strategic partner
in its confrontation with the Soviet Union.
Whereas on the US side that perception already changed
in the 1990s it remained unchanged on the side of China. 
\qpar

Among Chinese leaders there was a great admiration
for the US. Out of numerous facts
one can mention the following illustrations%
\qfoot{More details can be found in Roehner (2017a) particularly
in chapter 4 entitled ``Our constituencies''.}%
.
\qbu Leading Chinese universities provided
scholarship funding to their best students to allow them
to study in American universities and
pay the high tuition costs. 
Naturally, a substantial
percentage of them (95\% in 1987, 69\% in 2007
and 21\% in 2017, according to the ``South China Morning Post''
of 8 December 2018)
remained in the US which means that China accepted
to loose
some of its most promising students%
\qfoot{It can be noted that the first batch of Chinese students was sent
to the United States in 1872 that is to say 
by the Qing Empire. Japan had also sent its students to western
countries, but it seems it was done more cleverly in the sense
that they came back to Japan and greatly contributed to
its scientific and technological advancement.}%
.
\qbu Whenever a section of the Communist
Party wanted to express its satisfaction to one of
its members,
a common reward was in the form of a study tour in
the US. In other words, the best agent for spreading US influence
in China was the Communist Party itself.
As an illustration, one can mention that
from 2002 to 2014 the so-called ``Amway Program''
brought more than 500 Chinese officials to
Harvard's John F. Kennedy School of Government in order to
study public management. They were called
``Amway Fellows''. Another similar program at the
Kennedy School
was the ``China's Leaders in Development Program''
set up on the American side by Antony Saich%
\qfoot{In an article entitled 
``Amway bankrolls Harvard course for Chinese
cadres'', 
Bloomberg news of 24 September 2013 reported that 
next to Saich's office there was a cartoon on the wall
which pictured the same Chinese official some 20 years apart.
In the first frame, dressed in a Mao suit, he was raising
his fist before a Chinese flag saying
``I staunchly oppose America's hegemony!'' 
In the second frame, he was addressed by a seating official
in the following terms: ``You are very patriotic.
We will send you to Harvard for training next year!''}%
.
Such programs were certainly an excellent 
method for creating a network of Chinese officials fluent in English,
and sincere admirers of the American way of life.
On the Chinese side such
programs were supervised by the ``China
Development Research Foundation'' under the control of 
the ``State Council'', the chief administrative authority in China.
\qpar

In short, the United States was seen by Chinese leaders
and people as a well-meaning elder brother.
Such a climate can probably explain why, despite clear signs of
waxing US antagonism 
in the 2010s (e.g. massive sales of weapons to Taiwan%
\qfoot{The implicit understanding of the agreement
of 1979 which led to diplomatic recognition of the PRC
was that such sales would gradually slow down.
Instead there was an acceleration.}%
,  
more and more visits of the Dalai Lama to the
White House) the official policy remained unchanged.
It takes time for younger siblings to rebel against
their big brother.
Even in late 2018, despite the trade war 
and a string of actions targeting Chinese telecom companies,
an article of ``People's Daily''
(the official publication of the Communist Party) of 
4 December 2018 was entitled%
\qfoot{To some extent, this policy may be ``tactical''
in the sense that as time is on the side of China, 
it is better to postpone any confrontation as
long as possible. Naturally, the same reason may convince
the US side to start it as soon as possible, yet at the
same time avoiding (thanks to its hegemonic position in
world media)
being perceived as the aggressor.}%
:
``Jointly promote a healthy, steady China-US relationship''.

\qA{What does Table 2 say about the Chinese challenge}

Coming back to Table 2,
in spite of several interrogation marks in the row
about China, the clear conclusion is that challengers,
including China, should be
eliminated. As a matter of fact,
through its mere existence and economic
expansion, China is a more serious challenger
than was the Soviet Union in the 1960s and 1970s.
The fact that it will not be tolerated
means that the tension between the two countries
will not decrease but rather increase. 
In the next section, we examine how this prediction
can be tested.

\qA{``Who will win?'', a question that we cannot answer.}

For the human mind it is a natural tendency to 
wish a {\it detailed} view of the future. In particular,
one would like to know who will win. Will the challenge
be suppressed as it was in almost all previous 
cases except Vietnam?
\qpar

It must be strongly emphasized that 
Table 2 only allows
us to say that the challenge will not be tolerated 
and will give rise to a confrontation.
This will certainly
be found unsatisfactory by many readers,
but to predict who will win is a much more difficult
question.
\qpar

Instead of restricting ourselves to the regularities
which can be derived from Table 2 it may be tempting
to indulge
into an exercise of political fiction. We will refrain
from doing that. 
Just to indicate why
let us consider the column ``Countries acting for US''.
One would of course wish to know which countries may be among
the ``several'' indicated in Table 2. On
account of recent events,
it may be tempting to predict that Japan will be among them
for indeed the re-armament 
program undertaken since 2015 by the
government of Prime Minister Abe suggests that Japan
will increase its military role in the Pacific.
However, this prediction would not be based on
the methodology of recurrent events but rather
on a short-term anthropomorphic argument. 
Would Yukio Hatoyama
come back to power the propect would change completely.
Moreover, contrary to what happened in 1904-1905, this time
Japon will not be able (and would certainly be unwilling)
to do the job alone. In this respect, it can be observed
that the size and economic system of China makes its
challenge more serious than any of the previous challenges
listed in Table 2.

\qA{Is the US a declining power?}

This question goes in the same direction as the ``Who will
win'' question but it seems easier to answer because it
is a more structural issue. It is not surprising that
different authors give different answers for the question
has several facets. Whether we talk about economy, finance,
foreign relations, scientific leadership or military domination,
the answer will not be the same. Just as illustrations
consider the following facts.
\qbu {\bf Economic facet}\quad
The only reliable way to compare two economies is
in terms of ``Purchasing Power Parity'' because otherwise
the comparison depends on the exchange rate%
\qfoot{Another way to make a comparison that is fairly
independent of exchange rates is through foreign
trade (at least if it is mostly carried out in
dollars). In 2009, the foreign trade of China surpassed
the foreign trade of the US.}%
.
\qpar

According to the World Bank, in 2017
China's GDP at Purchasing Power Parity  was
23.3 trillion USD
whereas the US GDP was only 19.4 trillion USD,
that's a difference of 20.1\%. 
Actually, the fact that China is already No 1
is well recognized by
some lucid western publications; see for instance
a Blomberg article published on 18 October 2017
and entitled: 
\qdec{``Who has the world's No 1 economy?
Not the US. By most measures, China has passed the US
and is pulling away''.}
\qpar

In other words, economically, 
with respect to China the US is certainly a declining
power but so are also Japan and the European Union.
\qbu {\bf Foreign relations facet}\quad 
Instead of embarking in a broad discussion, one can observe that
in the Vietnam War the US had 5 allies, namely Australia, New Zealand,
the Philippines, South Korea and Thailand whereas in 
the war in Afghanistan (2001--Dec 2014) the US has been leading
a coalition of 44 countries. For the occupation of Iraq in the wake of
the invasion of 2003 the US had 36 Allies.

\qA{The ways and means of the confrontation can be predicted}

What can be predicted are the ways and means of the
confrontation for these features do not change
much in the course of time. They
were essentially
the same in the Roman Empire, during the Cold War
and now. \qL
A non exhaustive list would include 
the following facets.
\qee{$ \hbox{W}_1 $} Targeting of people (e.g. scientists) suspected
of informing the opposite side.
\qee{$ \hbox{W}_2 $} Economic, industrial and financial warfare.
\qee{$ \hbox{W}_3 $} Effect of confrontation on separatism.
\qee{$ \hbox{W}_4 $} Information warfare.
\qee{$ \hbox{W}_5 $} Containment policy.
\qee{$ \hbox{W}_6 $} Arms race
\qpar

In the following sections we describe the first three items;
the last three will be left for a closer study in
a subsequent paper.
\qpar

For each of the issues in the previous list we are on
fairly solid ground because we can rely on a series
of previous cases; for instance, several cases of
industrial warfare are listed in Table 3.

\qI{Events indicative of a brewing confrontation}

Within five or ten  years it will become easy
to judge if the Chinese challenge has
indeed led to a serious confrontation.
If we do not wish to wait several years, we can try to
detect in recent events indications of a brewing
confrontation.

\qA{Targeting of scientists seen as security risks}

\qun{The Cold War precedent: targeting socialists and communists}
\qL
During the first Cold War US scientists who were Communists
or had some sympathy for socialist ideas were 
kept under close surveillance by the FBI and in several cases
were blacklisted. This
was true even for prominent scientists like David Bohm,
Albert Einstein or Robert Oppenheimer. Unlike Einstein, Bohm
and Oppenheimer were both born in the US.
Several other colleagues of Oppenheimer at Berkeley were
blacklisted, e.g. Ross Lomanitz, Philip Morrison, 
Steve Nelson, Frank
Oppenheimer (Robert's younger brother), his wife Jackie
Quann and Joseph Weinberg. 
\qpar
It should be noted that
none of these physicists was accused of spying for the
Soviet Union; it was only assumed that because of 
their {\it opinions} they could become security risks. 
\qpar
 
The case of Bohm can serve as an illustration. A close
collaborator of Einstein at Princeton, he was called to 
testify by the ``House Un-American Activities Committee'' (HUAC)
in May 1949%
\qfoot{Contrary to senator McCarthy himself, the
HUAC was active over several decades (see Roehner
2007, p. 147). However, as it
was not a tribunal it did not offer
the usual garantees given to defendants: most often they had
no counsel and it was not uncommon that
false witnesses were used against them 
(see Schrecker 1986, Rader 1979). The HUAC was only
one of several similar committees, e.g. one can mention
the Rapp-Coudert, 
Canwell, Tydings, Jenner, Nixon committees or the
``Senate Internal Security Subcommittee'' (SISS).}%
. 
Following the non-compliance attitude recommended
by Einstein himself, he refused to testify.
As a result, he was arrested in early 1950, 
indicted for contempt
of Congress, nevertheless freed on bail and eventually acquitted by
a federal district court in May 1951. However, in the meantime
Princeton University
had suspended him and refused to reinstate him.
He found a position at the university of Sao Paulo, but once
in Brazil he had to give up his US passport which prevented him 
from traveling. In 1955 he was allowed to move to Israel
where he spent two years before eventually joining
Bristol University in the UK.
\qpar

While Bohm did not collaborate with the HUAC, many others 
who were called to testify did.
For instance, in his testimony%
\qfoot{The proceedings of the public hearings (most hearings
were in fact {\it not} public) are available on line.}
of 25 February 1953, 
Mr. Robert Davis,
a teacher from Massachusetts, gave at least 15 names
of persons who were members of the Communist Party in 1938--1939.
That is why, just like an epidemic, this hysteria
spread wildly (in the sense that the geometric series $ 15^n $
increases very fast).
Incidentally, in those years
the Communist party was
supporting the ``New Deal'' policy of President Roosevelt.

\qun{Targeting scientists who have contacts in China}
\qL
Coming back to the case of China, a law was passed
by the US Congress in April 2011 which reads as follows (simplified form).
\qdec{None of the funds made available by this Act may be used for the
``National Aeronautics and Space Administration'' (NASA) or the ``Office of
Science and Technology Policy'' (OSTP) to develop a bilateral policy or contract
of any kind with China.}
\qpar
In other words through this law scientists working in the 
US, whether US citizens or not
and whether or not employed by NASA, were barred from any NASA
or OSTP funding whenever they had contacts with Chinese 
research institutions.
\qpar

In fact, the collaboration interdiction did not start in
2011 but already in 1998.
Following an investigation by a congressional commission 
led by Christopher Cox, there was an embargo
on US-Chinese cooperation in space. However, the
target of the law of 2011 was 
much broader than a prohibition of official NASA-China cooperation.
It barred cooperation not only with space science researchers but
with all Chinese researchers whatever their fields.
\qpar

The bill of 2011 was sponsored by Representative Frank Wolf who
was well-known for his anti-Chinese positions and speeches. 
However it would be a mistake to think that it was a personal
matter. When he retired in 2015 he
was replaced in the ``Science Subcommittee''
by John Culberson who,
like his predecessor, vowed to uphold the
embargo on space cooperation with China.
Actually, in the shadow of Wolf and Culberson one would expect
the Pentagon to be the key factor.
The fact that the US has an ongoing collaboration with
Russia but refuses to cooperate with 
China clearly suggests that it is China
that is now seen as {\it the } strategic opponent.

\qA{Espionage charges}

\qun{The Cold War precedent}
\qL
Arrests on the charge of spying for the Soviet Union
started right after the end of the war.
The defection of Igor Gouzenko on 6 September 1945 led
to the arrests of 39 persons in the US and Canada
on the charge of espionage for the Soviet Union.
Numerous others were to follow in subsequent years.
These arrests targeted employees and civil servants
of the State Department but also many scientists.
 
\qun{Chinese American scientists arrested on espionage charges}
\qL
On 21 December 2018
US Deputy Attorney General Rod Rosenstein declared that between 2012
and 2018 more than 90\% of the Department of Justice cases concerning
economic espionage involved China (Taipei Times 22 December 2018).
\qpar

On Wikipedia there is an article entitled
``List of Chinese spy cases in the United States''
which describes the cases of 32 scientists, mostly 
of Chinese origin, who were arrested on espionage
charges. Except for a few older ones,
most of the cases occurred in the period 2000-2018.
The list is not complete; as examples of missing cases
one can mention: Gwo Bao Min, Xiaodong Meng, 
Billy Yui Mak, Fuk Heung Li,  Xiaoqing Zheng.
\qpar

The most striking fact is that in about one third
of the cases the charges were dropped either completely
or drastically narrowed in a way tantamount to recognizing
that the accusation had collapsed%
\qfoot{The following persons can be mentioned in this
respect (more details about them can be obtained
through Internet key-word searches.): 
Xiafen Chen,
Michael Haehnel,
Bo Jiang,
Peter Lee,
Wen Ho Lee,
Katrina Leung,
Tai Mak,
Xiaoxing Xi,
Hua Jun Zhao.}%
.
It is of interest to understand the reason.
\qpar

Most of the persons arrested were Chinese Americans (either
born in the US or naturalized US citizens) working
in the US and maintaining contacts in China.
These contacts may have been just for the purpose
of scientific cooperation or in the intention of
creating Chinese start-ups. In both cases, this
kind of activity involved moving scientific information
from the US to China. If this information was
deemed proprietary information without even being
officially classified it was enough to motivate
an arrest.\qL
In most cases where the accusation had to be dropped
it was because the defense side could convince the 
prosecution (represented at state level by a district attorney
or a US attorney at federal level)
that the information that was passed on was in fact freely
available in scientific journals. Needless to say, for scientific
or technical matters such distinctions can be difficult to make.
\qpar

Although we know a number of cases in which the accusation
was dropped there may also be cases in which defenders were
wrongly sentenced to terms of several years. This 
can be understood by considering the case of
Los Alamos scientist Wen Ho Lee. The government brought
against him 59 charges, including 39 that each carried a life
sentence. Fortunately, soon after his arrest a
prominent law firm who had agreed to represent him sprang into
action. In addition, led by Lee's daughter,  
a network of people and organizations
started to mobilize public opinion on his behalf and contributed
to cover the cost of Lee's lawyers%
\qfoot{In his book (Lee 2002) Lee says that the cost was
of the order of one million dollars. He thanks the 
following organizations which supported him:
(i) ``Asian Law Caucus'', (ii) ``Chinese for Affirmative Action'',
(iii) ``Organization of Asian Americans''.}%
.
\qpar

The question of
what would have happened without such a support is a matter of
speculation, but it is clear that not all scientists who
were indicted
benefited from such an effective support. In addition it can be
observed that in such affairs the political climate plays
a role; in 1999 when Lee was arrested US-China relations
were certainly more friendly than 20 years later.

\qA{Cluster of industrial warfare cases}

In 2018 the Chinese telecom giant Huawei was targeted
by several rules issued by the US government. At first sight 
this may not seem of great significance regarding
a looming confrontation for indeed in the past
such methods have also been used against foreign companies
belonging to US allies. However, the Huawei case was
quite different from the previous ones. To see this point
more clearly let us again use the analytical methodology 
employed for hegemony challenges.
\qpar

This time the mechanism can be defined as follows.
\qdec{\it {\color{blue} Industrial warfare}\quad
How to use US federal rules to limit the
penetration of foreign companies into the US market.}
\qpar

A set of such cases is displayed in Table 3.
%
%%-----------------------------------------------
\begin{table}[htb]

\small
\centerline{\bf Table 3: Cases in which foreign competitors
were targeted on technical grounds}

\vskip 5mm
\hrule
\vskip 0.7mm
\hrule
\vskip 0.5mm
$$ \matrix{
\qtb
&\hbox{Year}  & \hbox{Product} \hfill & \hbox{Country} \hfill 
& \hbox{Company or brand}\hfill \cr
\noalign{\hrule}
\qth
1&1973 & \hbox{Supersonic airliner (Mach 2)} \hfill & \hbox{France-UK} \hfill 
& \hbox{Concorde} \hfill \cr
2&1982 & \hbox{Car} \hfill & \hbox{Germany} \hfill & \hbox{Audi/Volkswagen}
\hfill \cr
3&1990 & \hbox{Mineral water} \hfill & \hbox{France} \hfill &
\hbox{Perrier}  \hfill \cr
4&2010 & \hbox{Car} \hfill & \hbox{Japan} \hfill & \hbox{Toyota}
\hfill \cr
5&2015 & \hbox{Car} \hfill & \hbox{Germany} \hfill & \hbox{Volkswagen}
\hfill \cr
6&2016 & \hbox{Cell phone} \hfill & \hbox{South Korea} \hfill &
\hbox{Samsung} \hfill \cr
7&2018 \hbox{ (27 Mar.)} & \hbox{Cell phone} \hfill & \hbox{China} \hfill &
\hbox{Huawei} \hfill \cr
8&2018 \hbox{ (15 Nov.)} & \hbox{5th generation smartphones} \hfill &
\hbox{China} \hfill &
\hbox{Huawei} \hfill \cr
\qtb
9&2018 \hbox{ (24 Nov.)} & \hbox{Telecom equipment} \hfill &
\hbox{China} \hfill &
\hbox{Huawei} \hfill \cr
\noalign{\hrule}
} $$
\vskip 0.5mm
Notes: \qL
Audi is the luxury brand of Volkswagen.
\qbu  On 27 March 2018 it became known that
a planned partnership between Huawei and the US telecom company ATT
will not happen due to political pressure.  
Verizon had similarly dropped plans to sell Huawei smartphones.
Thus Huawei phones could be sold only through independent retailers
which represents a very small part of the US market.
\qbu On 15 November 2018 it became known that 
the US and Australia
have excluded Huawei from the 5th generation auction for mobile phones
and that the British and German governments were being
pressured to do the same.
\qbu On 24 November 2018 it became known that the
US government was trying to persuade key allies 
to avoid using Huawei telecom equipment. This concerned
particularly the countries hosting US military bases.
\qL 
In subsequent weeks, there were news in a number of
countries of arguments between
governments and intelligence agencies. The latter
claimed that Huawei was definitely a threat to 
national security whereas the governments observed
that previous investigations had not found any
problem. In order to understand such disagreements
one should recall that since the end of World War II,
in all countries which are
US allies, the intelligence agencies have a close
but highly asymmetrical 
relationship with US intelligence agencies.
In contrast, the governments were trying to keep open 
economic opportunities.
\qbu Most of the cases listed in the table gave rise
to criminal investigations due to accidents allegedly
resulting from technical defects.
\qbu We did not include in this table the sanctions against
the Chinese telecom company ZTE, nor the arrest in early
December 2018 of Huawei's chief financial officer because
these events were not based on technical reasons like
others in this table, but on political reasons,
namely trading with Iran (although later on the charges
were changed in order to make the US extradition request
more acceptable by Canada). 
We neither included cases of
companies submitted to increased duties as the result of a 
trade war. 
\qL
{\it Sources: \qL
1: NYT (11 Jan 1978); 3: NYT (10 Feb 1990);
7: https://www.cnet.com; 8: ``Taipei Times'' (15 Nov 2018);
9:  ``Taipei Times'' (24 Nov 2018).} 
\vskip 2mm
\hrule
\vskip 0.7mm
\hrule
\end{table}
%%-----------------------------------------------
%

The cases 1-6 show that there is always a combination of
three elements. (i) Technical reasons which at first sight 
seem reasonable. (ii) An amplification of this technical factors
beyond what is reasonable. (iii) An intervention 
of the US government.
The weighing and timing of these elements may change but
they are always present in one form or another.
\qpar

The Huawei cases are different from the others in several 
respects.
\qee{1} It is no longer a technical reason that is given
but instead a national security reason.
\qee{2} The enacted rules no longer
concern a specific product but instead {\it all} 
Huawei products. In other words it is the
company itself which is targeted.
\qee{3} The industrial warfare against Huawei is not
limited to the US but instead is extended to all US allies.
Worldwide there are 77 countries which host US bases.
\qee{4} From March to November 2018 there is an
intensification of the offensive.
\qpar

It seems that Huawei
must be eliminated because its very existence and success
is an intolerable challenge to US technological hegemony
(just as was the Concorde some decades ago).
In short, the anti-Huawei campaign can be seen as a first
step in a coming confrontation with China.

\qA{Outcry in the US against Confucius Institutes}

When the Confucius Institutes were set up in 2002, their main purpose
was to teach the Chinese language. In this sense they were
similar to the ``Goethe Institutes'' for learning German
or the ``Alliance Fran\c{c}aise'' centers for learning French.
However, the Confucius Institutes were a more ambitious 
project in two respects. (i) Each foreign institute is located
on the campus of an American University. This contrasts with
the German and French institutes which have their own
buildings off campus. (ii) Each American university which hosts
a Confucius Institute has a partnership with a Chinese University.
\qpar

To establish partnerships with foreign universities is 
something fairly common in many countries worldwide.
However, the Confucius Institutes mixes two fairly different
traditions: (i) The Goethe-Alliance Fran\c{c}aise tradition
through which German or French culture is promoted abroad
and funded by the respective governments.
(ii) International partnerships in higher education
which is supposed to be funded by the respective universities.
When such partnerships are established with state
universities (as are the majority of German or French
universities) they are of course also indirectly
funded by the state but in a less visible way.
\qpar

In the present trade war climate 
one can hardly be surprised by manifestations of
hostility directed at Confucius Institutes. 
On the website ``The Hill'' (which reports
news from the US Congress) one can read
an article of 22 February 2018 which has the following title:
``Get China's pernicious Confucius Institutes out of US colleges''
and which ends with the following sentence:
``Confucius Institutes are an affront to intellectual freedom, national
security, and American interests. It is time for them to close, and it
is time for the US to act.'' \qL
For the time being (December 2018),
only four US universities in a total of about one hundred
hosting  Confucius Institutes have followed this advice.

\qA{Effects of China-US antagonism in South East Asia}

During the Cold War there was a power struggle in many
countries between leftist parties often suspected of 
sympathy for the Soviet Union and conservative parties
supported by the US. 
Leaders considered too friendly to the Soviet
Union were usually removed through a coup led by the army. Examples
are Chile in 1973, Thailand in 1976 or Pakistan in 1977.
One would not be surprised to see a similar effect as a result
of the China-US power struggle. This prediction seems
indeed confirmed by current events in several countries
of South-East Asia. However it would take too long to
analyze these cases here. One must realize that in each
country the antagonism between China and the US will
materialize in a different form determined by
domestic factors. This makes it difficult to decode
local events but at the same time knowledge of
the China-US antagonism gives a 
useful interpretation key.

\qA{Effects of China-US antagonism in Taiwan, Tibet and Xinjiang}

\qun{Core interests targeting as a thermometer of China-US 
antagonism}
\qL
Taiwan, Tibet and Xinjiang are usually referred to by the
Chinese Ministry of Foreign Affairs as part of China's
``core interests''. Therefore, the way they are
targeted by the US State Department reflects fairly well
the condition of China-US relations. For instance between 1980
and 1990, the period of strategic partnership,
there was not a single meeting between a US
president and the Dalai Lama%
\qfoot{The source is a Wikipedia article which lists
all trips of the Dalai Lama out of India. This article
also reveals that, in the same decade 1980--1990,
and in contrast with the US, there were 
four meetings of Pope John Paul II with the Dalai Lama.}% 
.

\qun{The case of Taiwan}
\qL
Taiwan-US relations reflect China-US antagonism in a
much ``cleaner'' way (that is to say with less noise
due to local factors) than US relations with 
neighboring countries.
\qpar

When one compares the so-called ``Six Assurances'' given
to Taiwan in 1982 by President Reagan with
the ``Taiwan Security Enhancement Act'' of 2000 (a bill 
never passed), the ``Taiwan Security Act of 2017'' 
(a bill in committee discussion since November 2017),
the ``Taiwan Travel Act'' (signed into law in March 2018),
one gets the feeling of an accelerated process
through which the US encourages Taiwan's independence%
\qfoot{ For instance in August 2018 Taiwan's president Tsai
was allowed to give a public speech in the US and to
visit NASA's Houston Space Center, a place closed to Chinese
space scientists.}%
. 
\qpar

In our comments about Table 2 we observed that US military
interventions were always%
\qfoot{The invasion of Iraq of 2003 seems to be the only exception.}
in response to an agression 
(e.g. Japanese attack of 1941, attack of 11 September 2001) or to
come to the help of a small country under attack 
(e.g. South Korea, South Vietnam, Kuweit). 
Pushing China to an intervention in Taiwan and then coming to 
Taiwan's rescue might seem an appropriate way to start
a limited conflict with China; but would it remain limited?

\qun{The case of Tibet}
\qL
The ``Reciprocal Access to Tibet Act''
was passed by the US Congress on 11 December 2018.
It requires the
US Secretary of State, within 90 days of the bill being signed into
law by the president,
to identify Chinese officials responsible for excluding
US citizens (particularly journalists)
from Tibet and then ban them from entering the United States.

\qun{The case of Xinjiang}
\qL
In recent months there have been US decisions targeting
the Chinese officials in charge of the administration 
of Xinjiang.
\qpar

Needless to say, all these actions are of a {\it political}
nature.
They show that the trade war which started in the Spring of 2018
is only one aspect of a global struggle.

\qA{Changes in number of Chinese graduate students in  the US}

So far most of our predictions were of the no/yes type,
that is to say we predicted the occurrence of some
new events and features. By putting together data given previously
we can venture to predict that from 2017 onward
the number of Chinese graduate students in the US will
grow at an annual rate which will certainly remain under 3\%.
Actually, one would, with good likelihood, expect it
to become negative. 
This prediction
is based on the combination of two factors.
\qee{1}  In the past 10 years the proportion of Chinese
students staying in the US after graduation
has shrunk from 69\% in 
2007 to 21\% in 2017. This limited the brain drain and
in addition when returning to China, these students brought
back the skills and knowledge acquired during
their studies. In a climate of
technological warfare such a feature will not be
seen with favor by the US government.
\qee{2} From 2007 to 2013 the number of graduate students 
in the US has increased at an annual rate of 20\%, but
from 2013 to 2017 the average increase rate fell to 3\%
(Mervis 2018). 
The inflection following 2013 was likely due to 
rising tuition costs combined with reduced Chinese
scholarships. 
\qpar

As one does not expect 
these trends to be reversed, the increase
rate should decline under 3\%, perhaps even become negative. 
As a matter of fact, it is 
this last case which would best agree with 
the expectation of Table 2.
\qpar

\qA{Financial implications of strained China-US relations}

Currently (in 2018) there are three strong financial
connections berween the PRC and the US.
\qee{i} China owns a sustantial part (over one trillion dollars)
of the foreign debt of the US federal government.
\qee{ii} More than one hundred major Chinese companies are listed
on the New York Stock Exchange.
\qee{iii} In 2018 about 80\% of the foreign trade of China was still
in dollars. 
\qpar

It is difficult to say what will happen in case of
an open confrontation for such a situation never happened before.
If financial exchanges become frozen one may expect the following
consequences.
\qee{1} US interest payments on Treasury bonds are likely to
be suspended. If one assumes an average
interest rate of 5\% this represents an annual amount of
\$50 billion, i.e. 0.40\% of the Chinese GDP of 2017.
In other words this effect is quite negligible. 
\qee{2} Chinese companies whose physical assets are located 
in China will probably interrupt dividends payments 
to American share holders. However, assets of Chinese companies 
located in the US or in allied countries may be taken over.
\qee{3} Interruption of the flow of dollars to China and
its allies will probably accelerate the dollar to renminbi 
substitution, at least in those countries.  
\qpar

Before coming to the conclusion we wish to
attract the attention of the reader on the 
discussion of the question of replication
which is given on Appendix C. This is
a key-point in our approach for the obvious reason
that replicability is the hallmark of science.

\qI{Conclusion}

\qA{Main results}

In this paper we have defined a methodology
for the scientific analysis of recurrent events
and we have used it in the investigation of
the relations between China and the United States
and also in the analysis of industrial warfare episodes.
This methodology was already presented (although
in a less formalized form) in a book
published some 16 years ago (Roehner and Syme 2002).
In the meanwhile its relevance and effectiveness 
were tested by using it in the investigation of
various phenomena (see Roehner 2007).
\qpar

Our main conclusion based on similar previous episodes
(listed in Table 2)
is that the very existence and development of China
is to be perceived as a threat by the US and will not
be tolerated. This means that the confrontation
will become harsher; it will take many forms
just as was the case during the Cold War.
We listed and described some events which are early
confirmations of this prediction.
\qpar

Actually, Table 2 and the ensuing conclusions
were already published in 
Di et al. (2017) and in Roehner (2017b),
that is to say prior to
the start of the trade war. At that time the bywords
were still cooperation and win-win relations. 
It is quite possible that, in line with the
wishes of big US corporations, 
there will be a truce in the trade war. 
However, this will not end the confrontation
in other fields.

\qA{A comment about the rationale for hegemony}

Some readers may observe that in this paper
we did not explain why a country should wish
to get and keep hegemony. It was by purpose and
we had mainly three reasons in mind.
\qee{1} The motivations may be quite diverse.
At the end of
World War II the Soviet Union wished to establish
its hegemony over East European countries mainly
for military reasons. As another example, in the 1950s
France wished to maintain
its hegemony in North Africa because there were
over one million French settlers in Algeria and these
people controlled a very effective lobbying network.
Even inside the same country different groups
may have specific motivations: the army will
insist on overseas bases, large corporations
want low taxes both domestically and abroad,
and so on.   
\qee{2} It seems fairly clear that through
its hegemony in India Britain secured economic benefits.
However, even in such an obvious case,
it would be difficult to give a reliable estimate.
In terms of annual GDP growth did they represent
0.1\%, 0.5\%, 2\% or more? If we cannot
estimate the benefits reliably is it worthwhile to  
describe them in detail. Moreover, as already said,
for the Pentagon GDP growth may not the main
concern.
\qee{3} Finally, the last and probably
most compelling reason results from 
our wish to follow the guiding 
lines of physics.
Physicists focus on the ``how?'' question,
and ignore completely ``why?''  questions.
Why do rain drops fall? Neither newtonian mechanics
nor General Relativity gives an answer but
the two theories give very accurate predictions
about how they fall, e.g. speed and acceleration.
As a matter of fact, this focus on ``how'' is
exactly what Durkheim says when he recommends to
study social phenomena as ``things'' so as to
avoid anthropomorphic interference.
\qpar

In short, to see some countries willing to
fight to defend their hegemonic position is enough
for us. We do not wish to spend time on asking why
they are doing that.

\qA{Future investigations}

In this paper we have described the different facets
of the confrontation rather briefly. In fact, for each 
facet one needs to identify, collect and compare
a series of similar cases. We plan to do that in
subsequent papers. 
\qL
For instance, with respect to the question of separatism,
one should compare several cases in which
the US has supported separatist movements in
foreign countries. This can be traced back 
to the early 20th century with 
the independence of Panama followed by Ireland,
India, Indonesia, Ukraine, Croatia, Kosovo.

\qA{What means ``sidelining one of the contenders''?}

 In the abstract we used the somewhat cryptic wording that
``it is only through the sidelining of one of the contenders that the
confrontation will end''. Is it possible to be more specific?
\qpar

First, one should observe, that
strictly speaking, the cases in Table 2 do not 
really allow to do
that for indeed their outomes are fairly different. In 1905 the
setback suffered by Russia was mostly confined to the Far East.
On the contrary, in 1945 Japan suffered a crushing defeat
which had long-term implications up to present-day.
Nonetheless, based on reasonable assumptions,
is it possible to be somewhat more specific about the
possible meanings of ``sidelining? 
\qpar

For the United States the answer is fairly simple in the sense that
having to share world hegemony with China would constitute
a significant set-back.
\qpar

What would be a significant setback for China is less clear
for, contrary to the USSR, China does not yet have a superpower
status that it is at risk of losing. However, the following
outcomes may be seen as significant US ``victories''.
\qee{1} To be able to
maintain a technological edge in critical hightech sectors
(such as the semiconductor industry) which are essential elements
of the supply chain%
\qfoot{In contrast, high speed trains or nuclear power plants
are not critical elements of the supply chain which is why
the US accepted to lose its leadership in such sectors.}%
. 
Although, as already mentioned, the 2017 PPP GDP 
(Purchasing Power Parity Gross Domestic Product of 2017) of China
already surpassed the one of the US, this would imply a
fateful enduring weakness.
\qee{2} Through clever use of softpower, 
(a combination of financial threats and sanctions, 
use of captive media, public relations
campaigns, military cooperation)
to be able to isolate China particularly from
its Asian neighbors (e.g. India, Vietnam, Malaysia, Indonesia)
or from the Central Asian countries located on the silk road. 
\qpar

This is but a partial list of possible developments.
It is by purpose that we mentioned neither military aspects
nor possible domestic upheavals 
for in such areas predictions are notably hazardous.

\qA{What can be learned from the case of the USSR?}

We already said that we will refrain from politics fiction
and only rely on series of recurrent events. One former occurrence
is the demise of the USSR. What can it tell us?
\qpar
If one discards the occurrence of a war, what else can make the US accept a
lower status? Although
the case of the USSR is certainly not the best possible parallel let
us nevertheless see what it suggests.
\qpar

Around 1975 after its victory in the Vietnam war the USSR was a very 
successful superpower. 
In East Asia
it had four vassal states, namely North Korea, Vietnam, Cambodia 
(at least after the occupation by Vietnam)
and Afghanistan. With its allies
it created the ``Council for Mutual Economic Cooperation'' 
(Comecon) which also
included East European nations.
However, the Comecon could not deliver much in terms 
of economic achievements despite the fact that the tripling of
the price of oil was a huge bonanza for the USSR which is a major
exporter.
\qpar

Then, at the end
of 1979 the USSR occupied Afghanistan which turned out to be a
disaster, and in 1989, 
encouraged by the US, the East European countries and even Ukraine,
Belarus and other Soviet Republics sided with the west.
Among other things,
they were of course expecting faster economic growth. Did that materialize?
\qpar

One can imagine a similar scenario.
Presently, the US has bases in 77 countries which are all more or less
vassal states.
What will happen if by establishing commercial links with China these
countries are led to turn their back on the US?
\qpar

We do not suggest that this scenario is likely.
In fact, under present conditions it {\it seems} very unlikely.
However, this is a purely anthropomorphic judgement, that is to say
of the kind Emile Durkheim cautioned us to avoid.

\appendix

\qI{Appendix A: Built up of US hegemony in the Pacific}

As the annexation of Guam,
Hawaii and the Philippines occurred only in 1898, one might
think that there was no real US interest for the Pacific until
the end of the 19th century. Thus, our
argumentation will involve two steps.
\qbu First, we explain that US naval and diplomatic activity
in the Pacific and in East Asia has already started
in the early and mid-19th century, that is to say
decades earlier than is usually assumed.
\qbu Secondly, we show that at the end of the 19th century
the US was already far ahead of all its contenders in the
Pacific in terms of national product and naval power.

\qA{Early inroads in the 19th century}

There are several facts 
which suggest that US involvement in the Pacific started in
the early-19th century and developed considerably in the
first half of this century.
\qbu  Diplomatic relations between the US and Thailand
were established as early as 1818; at that time Thailand was 
the only country
of South East Asia which was not part of the British Empire.
Then, in 1833 the two countries signed a ``Treaty of Amity and
Commerce''.\qL
In 1853 through the use of gunboat diplomacy, the United States
obtained the opening of Japan to western trade.
\qbu Despite the fact that annexation occurred only
in 1898, from 1840 on the Hawaiian monarchy was virtually
controlled by US missionaries who were so to say the
Trojan Horse of the State Department. For instance,
in 1874 following riots
the United States landed troops to restore order.\qL
The same can be said of Korea. Although
a ``Treaty of Peace, Amity, Commerce and Navigation'' was
only signed with Korea in 1882
the influence of US missionaries started several decades
before.
\qbu
In 1856, the US Congress passed the ``Guano Island Act''
which authorized US citizens to
take possession of unclaimed islands containing guano 
deposits, that is to say accumulated excrement of seabirds.
Guano was used for making gunpowder and fertilizer. In the 
course of the following decades at least 46 islands came
under US possession. They extended all the way from the coast of Peru and
Chile to the Philippines and
Indonesia; nine of them are still officially US Territories.
%
%%-----------------------------------------------
%%%% GUANO ISLANDS
\begin{figure}[htb]
\centerline{\psfig{width=17cm,figure=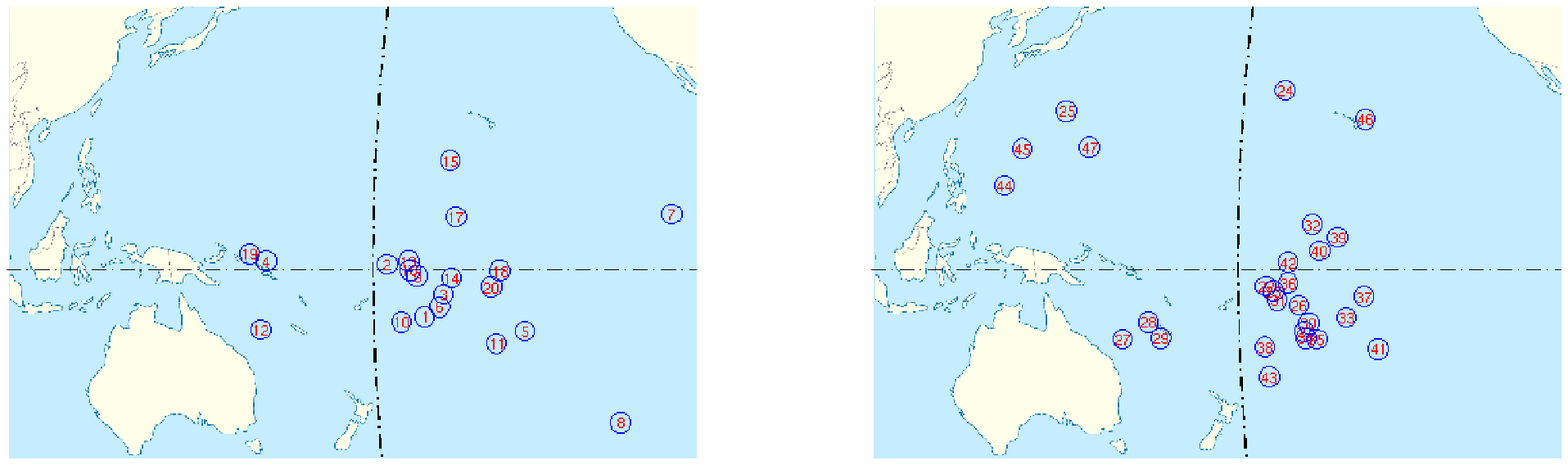}}
\vskip -15mm
\qleg{Fig.\qhu A1\qhv American guano islands.}
{The two pictures were separated for the sake of clarity.
Altogether they show 46 islands.
These islands were
occupied by the United States following the ``Guano Islands Act'' of
1856. Although many were occupied only temporarily,
others are still unincorporated US territories.
The two broken lines show the equator and the division line
between east (left) and west (right) of Greenwich.}
{Source: Wikipedia article entitled ``List of Guano Island claims''}
\end{figure}
%-------------------------------------------------
%
\qA{US hegemony around 1900}

The clearest indication of a watershed was of course 
the Spanish American War
of 1898 through which Guam and the Philippine islands 
were annexed by the US. A display of 
the effectiveness of US naval forces was given in the
Battle of Manilla Bay where
the Spanish squadron was destroyed within a few hours
with only 9 wounded on the American side.
\qpar

What was the situation of the other contenders?
At that time, the Chinese Empire was in chaos.\qL
Was Japan a more serious competitor? 
It is true that in the Sino-Japanese
War Japan won a resounding naval victory. However, one should
remember that around 1900 most of the Japanese warships were
still imported from Europe. At that time Japan had only 
a small industrial sector which weighed less than the 
agricultural sector and represented 24\% of national income.
In 1900 the US GDP was about 15 times the GDP of Japan.
In 1940,
the GDP ratio was still about 10 which means that in a war of
long duration Japan was no match for the US. This was 
of course even more true in 1900%
\qfoot{The detailed figures are as follows: US GNP=\$19,000 million,
Japanese income=2520 million yen, exchange rate: 1US\$=2 yen.
A comparison based on trade data leads to a ratio which is 
somewhat lower but one must take into account the fact that the
larger a country, the smaller its GDP/trade ratio. 
The sources are: Hundred-year statistics of the Japanese Economy,
Statistics Department of the Bank of Japan 1966, p.28 and
Liesner (1989, p.74,102,270).}%
.
\qL
The fact that the European powers (mainly Britain, France and Germany)
did not wish to challenge US hegemony in the central Pacific is well
shown by the fact that the annexation of Hawaii was hardly
opposed.

\qI{Appendix B: Threats to US interests in the Pacific}

\qA{Russian expansion toward Mongolia, Manchuria and Korea}

First of all, one should emphasize that the sale of Alaska to the US
in 1867 does not signal a lack of interest for the Far East.
Why? The main reason of the sale was because Russia realized it would
be unable to defend Alaska in case of a conflict with Canada
(still a British dominion) or the US. Secondly, Russian penetration
in Alaska was limited to a few fur traders. Thirdly, Russian
expansion was directed toward the southeast of the Pacific and for that
purpose Alaska was completely out of the way.
\qpar

In 1900 Outer Mongolia was still part of China
but because of the weakness of the Qing Empire it was in fact
controlled by Russia. From Outer Mongolia, Russian
influence could spread to Inner Mongolia and from
there to Tibet because of the strong cultural ties
between the two countries. It was partly in order to
prevent the spread of Russian influence that Britain
invaded Afghanistan in 1978--1880%
\qfoot{It can be noted that in its attempt
to fight off the invasion the Afghan leader asked for
Russian help.}
and Tibet in 1904--1905.
\qpar

Some of the landmark steps in Russian expansion can be summarized
as follows (detailed explanations can be found on Internet) 
\qee{1} 1860: Through the treaty of Peking, Russia got
Vladivostok; yet it was {\it not} an ice-free port.
\qee{2} 1875: Through the Treaty of Saint Petersburgh, Russia
received the Sakhalin peninsula.
\qee{3} 1896: Through the treaty Li-Lobanov, Russia was allowed 
occupation and administration of the Liaodong Peninsula including the
ice-free port of Port Arthur (now Dalian).
\qee{4} 1900: After the Boxer Rebellion Russia occupied the whole of
Manchuria with a substantial force.
\qee{5} 1901: The ``North Manchuria Railway'' was established
by a Russian company.
\qpar

Simultaneously, and in contrast to Alaska, there was a
substantial inflow of Russian populations. Its legacy is still
visible in the architecture of cities like Harbin and Dalian.
\qpar

Russian expansion was all the more perceived as a threat
because the US could do little to counter it.
Manchuria may have been far away from Moscow but it was even
more far away from Washington. At that time the US had no bases
in South Korea or Japan. Thus, in order to stop
Russian expansion the US administration had to convince Japan to 
confront Russia.

\qA{Japan pushed to wage war against Russia}

In the cartoon of Fig. B1 Britain is shown as
trying to interpose itself; it is indeed true
that Britain was much less pushing to war than the US
but nevertheless, after the war had started, Britain
was on the side of Japan and welcomed Japan's
victory%
\qfoot{In contrast France was a close ally
of Russia and tried to help it as far as possible
for instance by permitting a refueling stop
of the Russian Baltic Fleet in
a port of French Indochina.}
. 

%
%%-----------------------------------------------
%%%% Cartoon3
\begin{figure}[htb]
\centerline{\psfig{width=8cm,figure=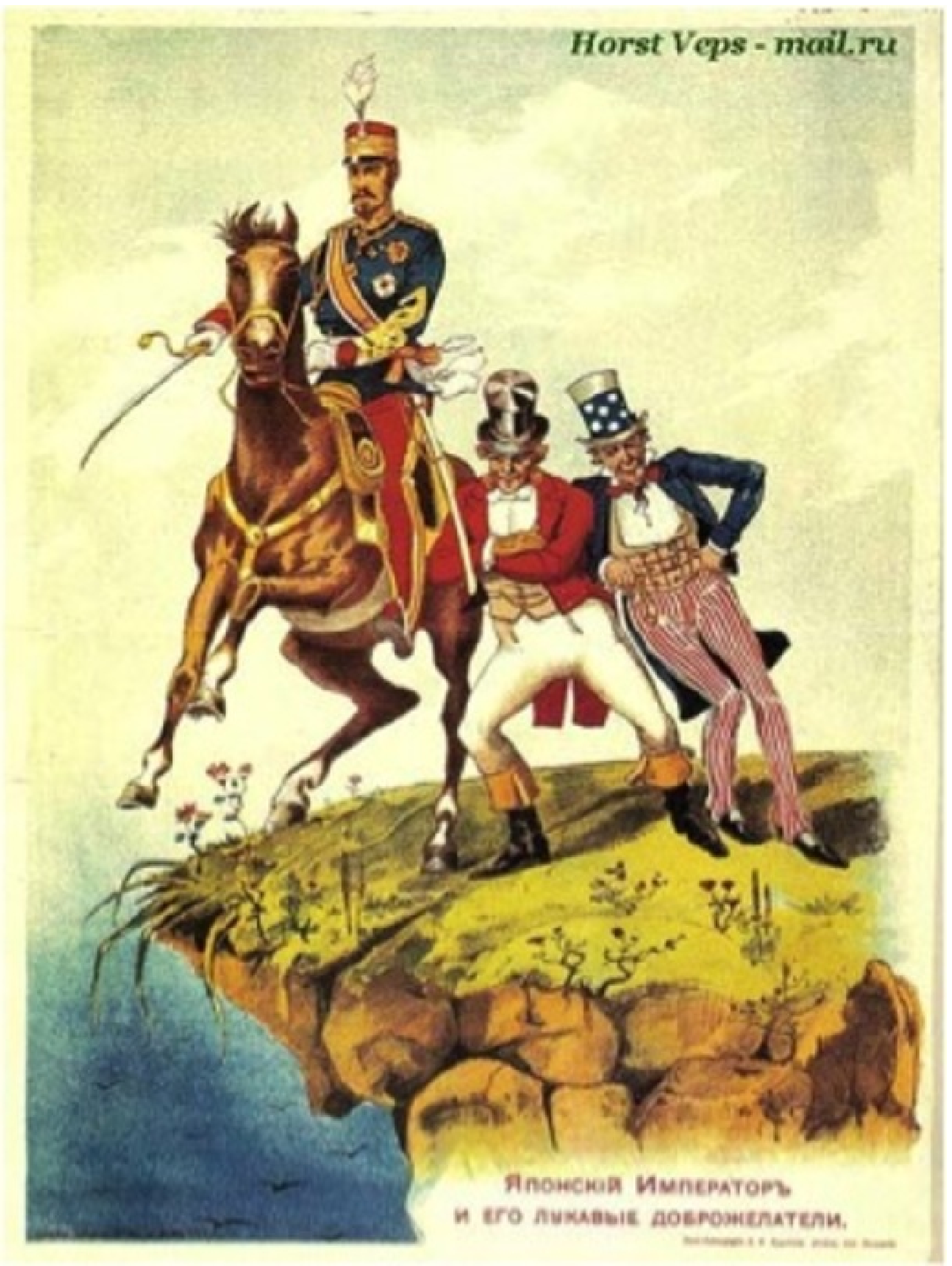}}
\qleg{Fig.\qhu B1\qhv Cartoon about the respective
roles of Britain and the US in the conflict
between Japan and Russia.}
{In this Russian cartoon
the USA is pushing the horse of the emperor of Japan
over the brink while Britain (in red) is
trying to interpose itself. At that time Britain had an
alliance with France which itself was a close ally of Russia.
In addition in 1902 England had signed a treaty with Japan.
Thus, it was certainly not the wish of Britain
to see Russia go to war against Japan.}
{Source: Internet}
\end{figure}
%-------------------------------------------------

From 
the articles published in the ``New York Times'' in the years 1895--1904
it is clear that this was indeed the policy of the State Department.
Whereas, Britain tried to encourage discussions between Japan
and Russia, the NYT seized every occasion, not matter how small,
to announce that the war was imminent. This was very clear for 
the public opinion as shown by cartoons of the time such as the
one shown in Fig.B1. 
In order to illustrate that the US played the role of a rablerouser
here are a few excerpts of the ``New York Times''. Usually the NYT
represents fairly well the positions of the State Department.
{\small
\qbu The Japanese Army is full of Russian spies disguised 
as Japanese. (24 Dec 1897) [A rather surprising allegation.]
\qbu Seventeen British warships back an ultimatum against
Russian domination. Japan supports the British action. Her fleet
of 30 vessels is awaiting the result of the protest
against the dismissal [in Korea] of an English customs
officer. (27 Dec 1897) [To see three major fleets go to war
for the dismissal of a customs officer would be quite
surprising. Indeed, nothing happened.]
\qbu Japan feels warlike. Captain Sakuzzi, who is in San Francisco,
says Russia is treating his country shamfully.
(1 Jan 1898) [Should one take great account of the declaration
of a captain?]
\qbu  The Korea question is settled. England, Japan and
Russia have arrived at an agreement. (10 Jan 1898)
\qbu Japan is prepared for war. (22 Jan 1898)
\qbu There is a war feeling in Japan. (6 April 1898)\qL
[Coming 6 years before the war and just weeks after a partial
agreement was reached such titles show that, contrary to
Britain, the US did not welcome a peace agreement.]
}
\qpar
More excerpts of that kind can be found in Roehner (2017, chapter 3).
\qpar

These excerpts also show that it was Korea which was
at stake. Japanese goals began to include Manchuria
only after 1917 when the grip of Russia over this region
was weakened as a result of the revolution.
It is true that in the war of 1905 there was a battle
for Mukden (present day Shenyan) a city which located
in South Manchuria but in fact
near the border with Korea; its control was necessary
for the purpose of securing the occupation of Korea.

\qA{The United States welcomes Japan's victory}

If, as we argued, 
the US pushed Japan to confront Russia one expects
of course that it was happy with Japan's victory. This
is indeed confirmed by NYT excerpts in the days following
the great Japanese naval victory of Tsushima (27-28 May 1905)
\qpar

{\small
Togo and the men who helped him win his Trafalgar of the Far East
shared with the heroes who fought and died for this country in both
eulogy and applause at the memorial exercises held in Carnegie Hall
last night.
(31 May, 1905)
\qpar

Japan has now free hand on land and can drive Russia from the
Pacific coast of Asia (3 June, 1905)
\qpar

The Russian cruisers must sail or be interned.
By denying the request of [Russian] Admiral Enquist for
an opportunity to repair his fugitive Russian cruisers [including
injured sailors] at Manila President
Roosevelt laid down a doctrine that is new.
(6 June, 1905)
}
\qpar

The last excerpt regarding Russian cruisers which found refuge
in the Philippines is particularly harsh and shows
very little sympathy for Russia in Washington.

\qA{Indonesia and its powerful Communist party}

 The second case that requires some explanations is Indonesia.\qL
Before its eradication in 1965 after a military coup,
the Communist Party of Indonesia (PKI in Indonesian) was the 
largest non-ruling  (that is to say outside
the Soviet Union and China)
Communist party in the world.
It had been founded in 1915 by Dutch socialists (Indonesia was
at that time a Dutch colony). Although banned by the Dutch authorities,
it was able to survive underground until when Indonesia became independent 
around 1947. Because of the role it had played in the fight for
independence and against the Japanese,
duly authorized by President Sukarno its membership grew quickly
to the point of reaching some 3 millions in 1964.
In February 1957 there was a first coup attempt by the pro-US
fraction of the military which failed. A second coup was
staged in October 1965 which started (as later on in Chile)
with the assassinations of all pro-Sukarno top generals.
It was followed by a terrible repression which, according
to most sources, claimed of the order of 500,000 deaths.
As a result, the PKI was eradicated and President Sukarno was
replaced by President Suharto.
\qpar

Why were Sukarno and the PKI perceived as a major threat 
by the United States? \qL
Firstly, there is a preliminary question. 
Was the PKI close to the Soviet Union or to China? Despite
the existence in Indonesia of a large minority of ethnic Chinese,
the PKI had closer links with Moscow. Secondly, the case of 
Indonesia should not be considered in isolation. It was part
of an expansion of Soviet influence not only in Asia 
but also in the Middle East with Baathist (i.e. 
socialist) coups in Iraq and Syria.

\qA{Chile, 1970-1973}

The case of Chile is fairly similar in the sense that
a leftist government was toppled over through a military coup
in September 1973.
This case should also be seen in the broader context of the
USSR-US confrontation.  The 
Paris Peace Accords of 27 January 1973 had officially ended
direct US involvement in the Vietnam War. They created a ceasefire
between North Vietnam and South Vietnam and
allowed 200,000 Communist troops to remain in the south.
Thus, they were in fact an acknowledgment of US defeat
as indeed confirmed by the fall of Saigon
in April 1975. In a position of weakness the US would not
tolerate any other challenge. Apart from the military coup in Chile,
there were also similar ones in Pakistan and Thailand. 

\qA{Why was Cuba tolerated? In fact, it was not}

Finally, a word is required to explain the case of Cuba.
Although Cuba is in the Caribbean Sea not in the Pacific,
this is certainly also an area that the Americans wish to
see as an American lake. 
Why, then, did they tolerate 
the leftist regime at their doorstep?
\qpar

Firstly, one should observe that in fact they did {\it not}
tolerate it in the sense that the failed ``Bay of Pigs'' landing of
April 1961 and the 5-year rebellion (1959--1963)
in the
Escambray Mountains against the Castro government
were clear attempts to remove it.
However, the real turning point was the missile crisis of October 1962
for the agreement which settled the crisis implied that
together with Soviet missiles in Cuba, US missiles in Turkey would 
also be removed and that no other invasion attempt would be made
in Cuba. 
\qpar

In the decades following the missile crisis the United States
continued its attempts to remove the Castro government. As
open action was barred, it used some of the 
other standard and less visible
means: (i) Containment policy by having Cuba excluded from
the  ``Organization of American Countries''; the exclusion
was only lifted in 2009. (ii) Economic and financial embargo%
\qfoot{Every year since 1991 the US embargo is denounced
by the General Assembly of the United Nations.}%
;
however several US allies did not enforce the embargo.
(iii)  Support to organizations
Cuban defectors in Florida (e.g. the ``Cuban American
National Foundation'')
and to domestic opponents in Cuba. (iv) International public relations
campaigns directed against Cuba.
\qpar
In short, it is only thanks to Soviet support and cracks in the
anti-Cuban coalition that the socialist regime 
was narrowly able to survive. In 2015-2016 there was
a short-lived normalization
in which Pope Francis played a key-role but a 

\qI{Appendix C. Replication in history versus physics}

The question of replication has two facets.
\qbu Firstly, we want to specify under what conditions 
historical events can be separated from their context. 
\qbu As the questions of uniqueness and replication
are a key issue we want to explain why from physics to history
there is a continuity rather than a radical change.
This discussion could have been presented in the introduction
but it may better be understood
once a case-study has been presented.

\qA{Can one extract events from their historical context?}

At first sight the Russo-Japanese War and the Vietnam
war seem to have very little in common. As a matter of fact,
the time intervals as well as the countries
involved  are not at all the same. However, we do not intend
to compare them; the only thing in which we are interested is 
the fact that they took place. Similarly, from an anthropomorphic
perspective an apple and the Moon seem
completely different objects; what they have in common is
that they are both attracted by the Earth according to the
law of gravitation.
\qpar

The advice to refrain from studying social phenomena 
from an anthropomorphic perspective  
was not given by an econophysicist
but rather, more than one century ago,
by Emile Durkheim, one of the founding
fathers of sociology (Durkheim 1894).
In his book about the methodology of sociology 
he devotes many pages to this question
and insists on the fact that social effects should
be studied ``like things'' (in the English translation), 
``comme des choses'' in the original French text.
Clearly such ``things'' can be extracted from their
historical context for the purpose of comparison.
We wished to emphasize this point because it is 
often a source of misunderstandings with historians.

\qA{Replication in physics and in history}

\qun{Semi-repetition in history}

The fact that something which has happened several times
is likely to happen again is a crucial step in our
methodology but one must recognize that this statement
must be given a fairly ``elastic'' meaning. 
What we mean is best explained by a few examples.
\qbu If sunrise was at 6:00 three days ago, at 6:05
two days ago, at 6:10 yesterday and at 6:15 
today is it not likely to occur at 6:20 tomorrow?
This may seem a trivial example, but if we look at it more
closely it is less obvious than it could seem.
Firstly, one must assume that the person remains at the same
location. If he or she moves
eastward or westward the prediction will not hold.
Moreover, even if the person does not move, the prediction 
is only approximately correct. 
\qbu A more realistic illustration is provided by
influenza outbreaks. According to statistics for New York City
covering a period of 31 years from 1889 to 1919, the monthly
number of deaths due to influenza displayed a peak
usually in December-January. This regularity allows fairly accurate
forecasts, at least in ``normal'' years. However, it is well-known
that for reasons not yet well understood, in 1918 the
outbreak took place in early November and its magnitude
was some 8 times higher than in normal years. This example
shows that the predictions based on recurrent events are
probabilistic rather than deterministic. This is related to the 
fact that there are parameters which are either not well known
or not well controlled. This leads us to replace the notion
of recurrent events by the broader notion of {\it paronymous events.}
It is explained below.
\qbu Everybody knows what are homonyms; they are words which have the
same spelling but different meanings; an example is
left (past tense of the verb ``to leave'') and
left (opposite of right). The notion of paronyms is less
well known but is merely a generalization:
paronymous words are like
homonyms whose spellings are allowed to be slightly
different; examples are: collision-collusion, differ-defer,
continuous-contiguous. The main purpose of this change in
vocabulary is to acknowledge two things (i) That historical events 
do not, strictly speaking, repeat themselves. (ii) That we
are more interested in the form of the events than in their meaning. 
\qpar

Paronymous repetition is a weak form of replicability.
Does this introduce a drastic difference with what we see in
physics? Actually, in the next subsection we show that
even in physics, strictly speaking, it is
impossible to repeat an experiment.

\qA{Semi-replication in physics}

The main objection of historians to the comparative methodology
is to say that historical events are unique and therefore
cannot be compared. Here we show that the same objection can
be made to experiments in physics. 
To prove this point we do not need to consider a sophisticated
experiment in quantum physics. Instead, our argument is
based on one of the simplest possible experiments in classical
mechanics, namely the swing of a pendulum (Fig. 2).
\qpar
 
%
%%-----------------------------------------------
%%%% MESURE DE LA PERIODE DE 2 PENDULES 
\begin{figure}[htb]
\centerline{\psfig{width=14cm,figure=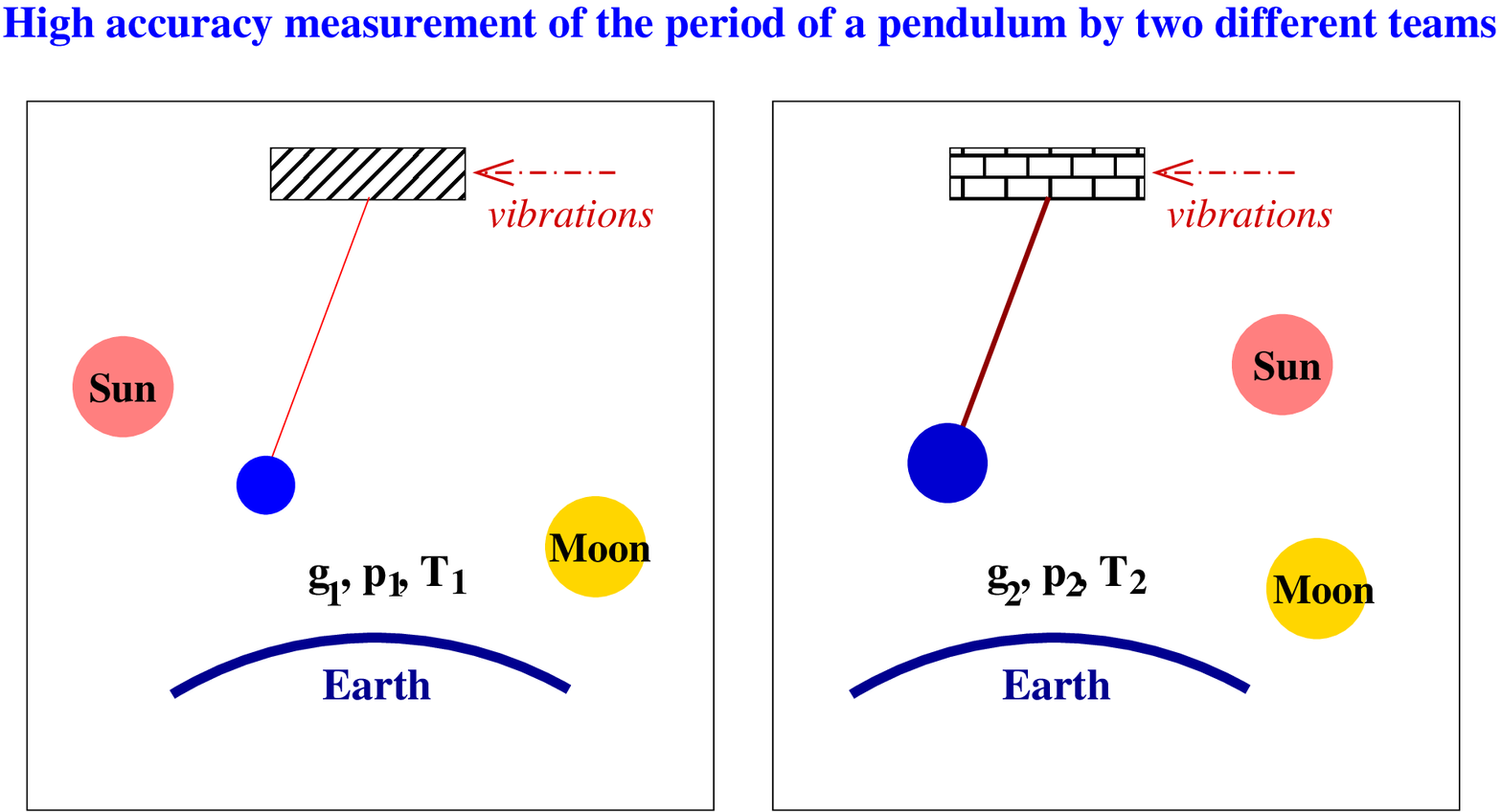}}
\qleg{Fig.\qhu C1a,b\qhv High accuracy measurements of the
periods of a pendulum.}
{When a high accuracy is desired one must take into account
many effects: amplitude of the angular deviation
(the small angle approximation will no longer be
sufficient),
local gravity, air pressure and temperature,
tidal effect due to the attraction of the Sun and Moon.
As a result, the experiment becomes strictly speaking
non reproducible in the sense that when performed on
successive days it leads to different results.
It is only when all such
changing perturbations
are well understood and can be corrected that
successive measurements (done either by the same team
or two different teams) become comparable.
For over three centuries physicists have been working with
patience and determination to get a full understanding 
of all such secondary effects.
From Galileo (circa 1600) to Friedrich Bessel (circa 1840),
to Thomas Mendenhall (circa 1920) the accuracy of
the measurement was improved. For instance,
in 1817 by using a new type of pendulum Henry Kater was able
to determine the length of a seconds pendulum (i.e.
a pendulum whose period is 2s) as equal to: 
$ L=994.137 \pm 0.003 $mm.
Such accurate measurements in turn led to 
several important discoveries
for instance the measurement of the 
density of the earth.}
{}
\end{figure}
%-------------------------------------------------

The key-point is the connection between replicability
and accuracy. 
\qbu If the period of the pendulum 
is to be measured with a low precision of one second,
then even two fairly different experiments in terms of 
location (which influences the gravity $ g $), diameter
and elasticity 
of the wire, air pressure and temperature 
will give the same results,
provided the two pendulums are (approximately) of same length.
\qbu On the contrary, is the period to be measured with millisecond
accuracy then all previous parameters must be controlled.
Moreover,
it is likely that due to differences in vibrations the
results on week days will differ from those obtained during
weekends.
\qbu If nanosecond accuracy is required then one must take into
account the tidal force due to the combined effects of the 
Sun and Moon. This makes the measurements
fully time-dependent, in the sense that two observations done
at 12:00 and 15:00 respectively will give different results.
\qpar

In summary, when only low accuracy is demanded even fairly
different historical episodes can be considered
as acceptable realizations of the same core mechanism. 
This is all the more true when the observations are
of a qualitative nature%
\qfoot{A qualitative description can be seen as 
made in terms of  ``no-yes'', i.e. $ 0-1 $, which corresponds to
error bars of the order of $ 1/0.5=200\% $.}%
.

\qI{Appendix D. A case-study book by political scientists}

\qA{A remarkable book}

In April 2018
a book%
\qfoot{It has 275 pages, with 61 pages of notes and index;
100\% of the references are sources in English.}
was published by Cornell University Press (MacDonald et
al. 2018) which investigates the issue of power struggles between
nations. It is written by two political scientists but
whereas most of the papers or books on the question
of China-US relations have a short-time perspective which hardly
exceeds two decades, this one relies on a series of case-studies
which goes back to 1870. A summary table (Table 1, p.22) which
lists 16 cases ressembles our Table 2 in the sense that it
lists both the challengers, the powers being challenged
and the outcomes. At first sight, this seems a sound
starting point and methodology. 

%
%%-----------------------------------------------
%%%% TABLEAU DE MACDONALD
\begin{figure}[htb]
\centerline{\psfig{width=12cm,figure=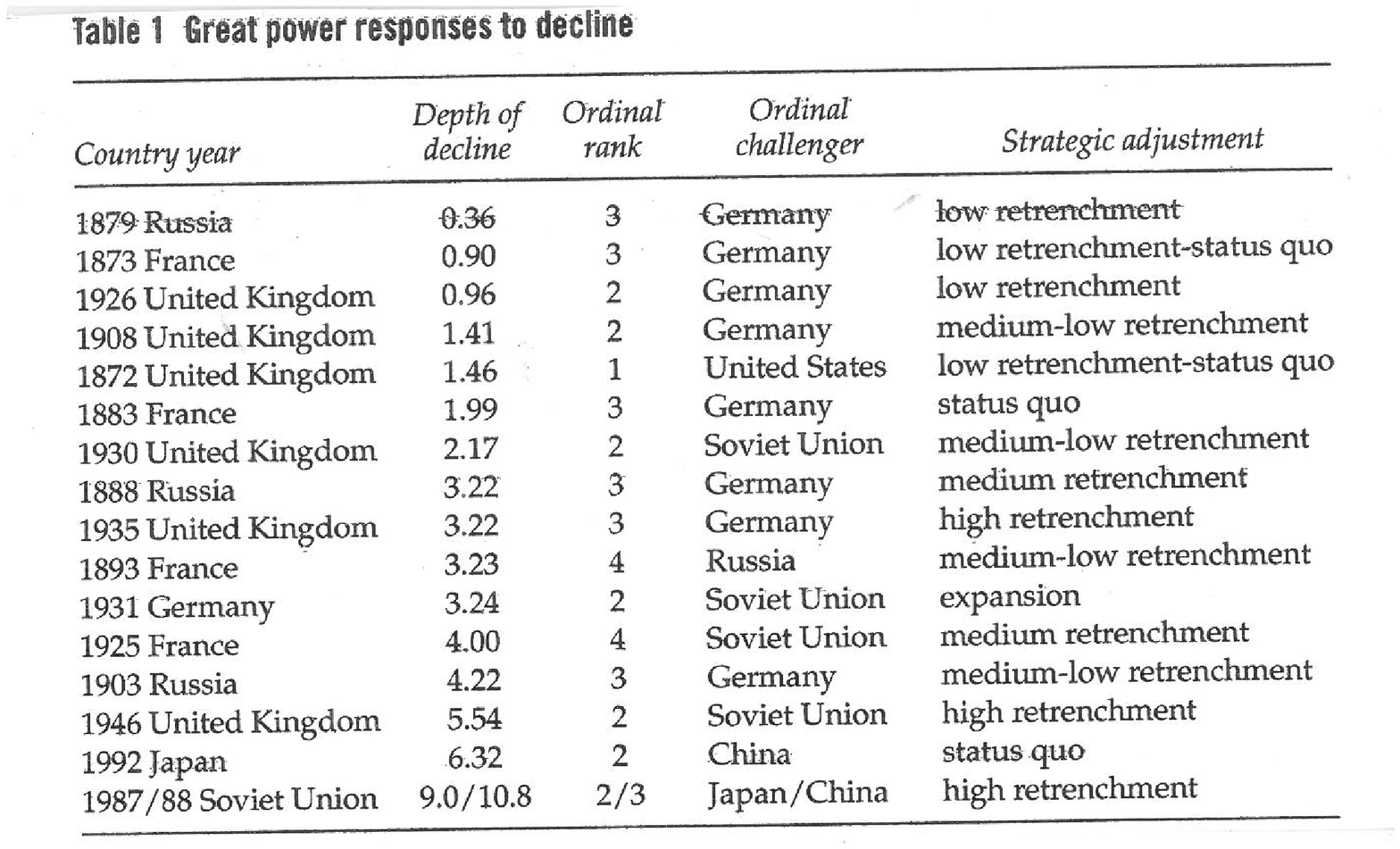}}
\qleg{Fig.\qhu D1\qhv Cases of responses to
national declines given in MacDonald et al. (2018).}
{``Ordinal rank'' means the rank of the country in terms
of world leadership.
In its spirit this table ressembles our Table 2 in
the sense that it lists similar realizations of the same
phenomenon. However, a crucial difference is that
these cases are small incidents
rather than major events as in Table 2.}
{Source: MacDonald et al.(2018, Table 1, p.22).}
\end{figure}
%-------------------------------------------------
Why, then, do the authors arrive to 
a conclusion which is the exact opposite of ours?
Indeed, in their conclusion they say:
\qdec{``If our
theory is correct, then the United States will respond
to its impending decline by retrenching.
Yet retrenchment will be mild and war very unlikely'' (p.191).}

\qA{A misguided comparative methodology}

In implementing their comparative program the authors neglect
three key lessons that experimental physics tells us.
\qbu The mechanism that one wishes to explore must
be clearly defined. In the 16 cases listed in Table 1
there does not seem to be a common mechanism. In some cases
a neighboring country tried to take advantage of the
supposed decline of a country, in others it is
the challenge itself which led to the decline.
In other words, this is very different from the
challenges to an hegemonic country listed in Table 2.
\qbu When one wishes to study a new phenomenon
(e.g. period of a pendulum or surface tension) one should 
design experiments which display this phenomenon with
greatest possible clarity and strength. One should not
try to observe it in cases where the effect is weak 
for in such cases it will be in competition with 
other effects which will make any conclusion uncertain.\qL
Regrettably, this is what the authors are doing.
In the langage of physics, instead of concentrating 
on first-order effects they mix these with 
second- and third-order effects.
The cases considered in their Table 1 are
not restricted  to major strategic challenges but includes many
situations of short-time regional challenges.
For instance, in 1883 there is mention of Germany challenging
France. However, in 1883 France did not have a hegemonic 
position in Europe. In 1908 there is another mention of Germany
challenging the UK, but curiously the German challenge of the war of 1914
directed against both France and the UK does not appear in the table%
\qfoot{There are other cases which are difficult
to understand. The United States is involved
in only one of the 16 cases and this is not the Japanese challenge which
led to the Pacific War, but a problem with the UK in 1872.
Quite surprising!}%
.
\qbu The second problem is that the authors do
not follow Emile Durkheim's advice to avoid anthropomorphism.
Instead of relying on hard facts and let them speak
by themselves, they depend on statements, opinions and motivations
of political leaders. This adds further uncertainty for
at any time one can find leaders advocating one course of
action or its opposite.
\qpar

In their conclusion chapter the authors rely mainly on declarations
made by President Obama and members of his administration
which suggest a policy of moderate retrenchment.
They overlook completely the aggressive sanction policy of the
Department of Justice or the restrictions to scientific
exchanges with China (see above). 
\qpar

Anyway, the best test will be to see what the future will tell us.
This book was published
shortly before the beginning of the trade war%
\qfoot{Let us recall that our Table 2 was first 
published in 2017.}%
.
In the meawhile there were also several
pro-Taiwan bills passed by Congress. All this
points to a direction which is at variance
with the predictions of the authors. Of course,
one should not rely on such short term events,
let us see what will happen in the next five
years.

%%%%%%%%%%%%%%%%%%%%%
%%% OMIS CAR ON A DEJA 2 EXEMPLES ET CELUI-CI N'APPORTE RIEN
%%% PARTICULIEREMENT EN FIN D'ARTICLE 
\count101=0  \ifnum\count101=1

\qA{Micro-historical events}

In ``Time Magazine'' of 17 December 1965 one can read
the following account in relation with
the events that followed the military coup:
\qdec{``Communists, red sympathizers and their families are being
  massacred by the thousands. The murder campaign became so brazen in
  parts of rural East Java, that Muslim bands placed the heads of
  victims on poles and paraded them through villages''.}
\qpar

We mention this behavior because in fact it is by no means
specific to this historical episode. It is one case in a long
series.
In Roehner (2004, p.56-57) we
give a list of 16 similar cases ranging from year 45 in
the Roman Empire to year 1987 in Saudi Arabia. One of the most well
known cases occurred in Paris on 14 July 1789 after the Bastille
castle was taken by the insurgents.
\qpar

Whereas the challenges in the Pacific were
macro-historical events, placing heads of opponents on poles or
spikes is a micro-historical event. Simple events of that kind should 
in principle be easier to analyze, at least if one can find
sources containing detailed accounts.

\fi
%%%%%%%%%%%%%%%%%%%%%%%

\vskip 3mm
{\bf Acknowledgments}\quad One of the authors (B.R.) would like to 
express his gratitude to the colleagues who gave him the
opportunity to give lectures or hold
discussions in their respective departments,
particularly to:
Hideaki Aoyama (Kyoto University),
Yuji Aruka (Chuo University),
Xiaosong Chen (Beijing Normal University),
Marco Cirelli (University of Paris-Sorbonne),
Zengru Di (Beijing Normal University),
Yoshi Fujiwara (Kobe-Riken),
Marc Gingold (CEA),
Beomjun Kim (Sung\-kyun\-kwan University),
Ruiqi Li (Beijing University of Chemical Technology),
Wei Pan (Peking University),
An Zeng (Beijing Normal University).
These lectures and discussions served as a testing ground for
the ideas and cases developed in the present paper.
\vskip 5mm

{\bf References}

\qparr
Baugh (R.F.), 
Archer (S.M.), 
Mitchell (R.B.), 
Rosenfeld (R.M.), 
Amin (R.), 
Burns (J.J.), 
Darrow (D.H.), 
Giordano (T.), 
Litman (R.S.), 
Li (K.K.), 
Mannix (M.E.), 
Schwartz (R.H.), 
Setzen (G.), 
Wald (E.R.), 
Wall (E.), 
Sandberg (G.), 
Patel (M.M.) 2011: Clinical practice guideline: tonsillectomy in
children. 
Otolaryngology-Head and Neck Surgery 144, 1 Suppl., S1–S30. 

\qparr
Carlson (S.) 1985: A double-blind test of astrology.
Nature 318,419-425.

\qparr
Di (Z.), Li (R.), Roehner (B.M.) 2017: USA-China: cooperation or 
confrontation. A case study in analytical history (in Chinese).
This book
is still unpublished but it can be read at the following address:\qL
http://www.lpthe.jussieu.fr/$ \sim $roehner/prch.php

\qparr
Durkheim (E.) 1894: Les r\`egles de la m\'ethode sociologique.
Flammarion, Paris.
[The book has been translated into English under the title:
``The rules of sociological method''. Both French and English
versions are freely available on Internet.]

\qparr
Ferguson (N.) 2010: Complexity and collapse. Empires on the
edge of chaos. Foreign Affairs 89,2,18-32.

\qparr
Lee (W.H.), Zia (H.) 2003: Me vs. my country.
Hachette Books.

\qparr
Liesner (T.) 1989: One hundred years of economic statistics.
Facts on File, New York.

\qparr
MacDonald (P.K.), Parent (J.M.) 2018: Twilight of the titans.
Great power decline and retrenchment.
Cornell University Press, Ithaca.

\qparr
Mervis (J.) 2018: More restrictive US policy on Chinese
graduate student visas raises alarm. Science 11 June 2018. 

\qparr
Rader (M.) 1979: False witness. University of Wshington Press.
Seattle.

\qparr
Roehner (B.M.), Syme (T.) 2002: Pattern and repertoire
in history. Harvard University Press, Cambridge (Mass.).\qL
[An updated version is available at:\qL
http://www.lpthe.jussieu.fr/$ \sim $roehner/prh.pdf]

\qparr
Roehner (B.) 2004: Coh\'esion sociale. Odile Jacob, Paris.

\qparr (B.M.) 2007: Driving forces in physical, biological,
and socio-economic phenomena.
Cambridge University Press, Cambridge (UK)

\qparr
Roehner (B.M.) 2017a: How China almost became an American
backyard. LPTHE Working Report, Paris.

\qparr
Roehner (B.M.) 2017b: USA-China: cooperation or confrontation?
A case-study in analytical history. LPTHE Working report.\qL
[Is available on the ``Wayback Machine'' website and on:\qL
http://www.lpthe.jussieu.fr/$ \sim $roehner/compet.pdf]

\qparr
Schrecker (E.W.) 1986: No ivory tower. McCarthyism and
the universities. Oxford University Press, New York.

\qparr
US House of Representatives 1953: 
Committee on
Un-American Activities, Communist methods of infiltration
(education). Public hearing of February 25, 1953, Wahington DC.\qL
[This publication is part of a massive set of similar accounts 
which totals 5,837 pages (available on Internet)]

\end{document}